\documentclass[printer]{aa}

\bibpunct{(}{)}{;}{a}{}{,}

\usepackage{graphicx}
\usepackage{txfonts}
\usepackage{soul}
\usepackage{txfonts}
\usepackage{color}
\usepackage{ulem}
\usepackage{lscape}
\usepackage{longtable}
\usepackage{rotating}
\usepackage{pdflscape}
\usepackage{subcaption}
\usepackage{xcolor}
\usepackage{threeparttable}

\newcommand{\hii}{H\textsc{ii}}
\def\ks{km s$^{-1}$}

\def\cm3{cm$^{-3}$}

\def\2{$^{12}$CO}
\def\3{$^{13}$CO}
\def\8{C$^{18}$O}

\def\cm2{cm$^{-2}$}

\begin{document}

\title{Sulfur-bearing molecules in a sample of active star-forming cores}

\author{R. D. Taboada\inst{1,2,4}
\and N. C. Martinez\inst{3,4}
\and S. Paron \inst{3}
\and A. Petriella\inst{3}
\and M. E. Ortega\inst{3}
\and L. Supán\inst{3} 
}
\institute{CONICET - Universidad Nacional de Salta. Instituto de Investigación en Energía no Convencional. Salta, Argentina\\
             \email{rocio.taboada@exa.unsa.edu.ar}
\and Universidad Nacional de Salta. Facultad de Ciencias Exactas. Departamento de Matemática
\and CONICET - Universidad de Buenos Aires. Instituto de Astronom\'{\i}a y F\'{\i}sica del Espacio. Buenos Aires, Argentina
\and Universidad de Buenos Aires. Facultad de Ciencias Exactas y Naturales. Departamento de Física. Buenos Aires, Argentina
}

\offprints{R. D. Taboada}

   \date{Received <date>; Accepted <date>}

\abstract{}{Astrochemical processes involving sulfur are not yet well understood because cosmic sulfur reservoirs and the production pathways of sulfur-bearing species remain elusive. Addressing this, requires high-resolution interferometric observations capable of probing individual molecular cores. Following our previous study focused on early molecular cores, we are motivated to expand this investigation toward a sample of evolved cores to understand their chemical transition.
}
{We analyzed data from the Atacama Large Millimeter Array toward 16 molecular cores in massive star-forming regions associated with methanol masers, targeting the same six sulfur-bearing species studied in a previous work toward early molecular cores: SO, SO$_2$, H$_2$CS, SO$^+$, NS, and $^{34}$SO. Column densities and molecular abundances were derived assuming local thermodynamic equilibrium, and kinetic temperatures were estimated from methanol transitions. Comparisons were made between the results obtained for the evolved cores and those previously obtained for the early ones.}
{We find that the abundances of all analyzed sulfur-bearing molecules are higher in the evolved cores than in the early ones, confirming a general time-dependent enrichment of sulfur in the gas phase. While abundances increase within the 100–220 K range, their correlation with temperature weakens, suggesting that gas kinematics become increasingly more important in the sulfur chemistry. This evolutionary transition involves a chemical reorganization where SO$_2$ becomes dominant, as evidenced by systematically higher abundance ratios such as SO$_2$/$X$. Based on our results we confirm the validity of SO$_2$/SO as a chemical clock, though chemical modeling reveals a discrepancy in sources with probably more pronounced kinematic processes,  namely a steeper increase in the SO$_2$/SO ratio. The line-width analysis of the molecular species indicates that core evolution and kinematics lead to a well-mixed gas, erasing the spatial stratification where different species trace distinct layers in the cores at earlier stages.}
{}

\titlerunning{Sulfur-bearing molecules in star-forming regions}
\authorrunning{R. D. Taboada et al.}


\maketitle

\section{Introduction}

Sulfur is a highly relevant element in astrochemistry, as its behaviour in the interstellar medium (ISM) remains one of the major open questions in the field. This is because its abundance in dense molecular regions is significantly depleted relative to diffuse clouds \citep{vastel18,laas19,riviere19,fuente23}. In fact, gas-phase sulfur abundances derived from molecular emission in dense ISM areas are estimated to account for, at most, a few percent of the cosmic value \citep{woods15,fontani23}. 

The main question, therefore, concerns the location of the `missing' sulfur in these interstellar regions. It is widely hypothesised that the missing fraction is locked into dust grains within the densest areas of the ISM \citep{laas19}. This trapping likely involves ice-mantle species, such as organo-sulfur molecules, sulfur chains, or rings like S$_{8}$ \citep{shing20}. However, OCS remains the only sulfur-bearing species detected in ices to date  \citep{geballe85,palumbo95}, with only upper limits established for H$_{2}$S,  \citep{smith91,jimenez11,hily22}, and tentative detections of solid SO$_{2}$ \citep{boogert97,zasowski09}. Thus, the primary sulfur carrier in the ISM is still unidentified. 

Despite extensive efforts to track the `missing' interstellar sulfur, its primary reservoir remains elusive in both the gas and solid phases \citep{artur23}. Consequently, further investigation into sulfur-bearing species within dense interstellar regions is essential to address this unresolved issue.

Using single-dish data, \citet{fontani23} investigated an illustrative sample of high-mass star-forming cores at different evolutionary stages, with five sources in each of the following categories: high-mass starless cores (HMSCs), high-mass protostellar objects (HMPOs), and ultra-compact \hii~regions (UC\hii). The authors found that the chemistry of sulfur-bearing molecules reveals a strong coupling between gas-phase molecular abundances and the evolutionary stage of the cores. They point out that during the transition from HMSCs to UC\hii, the incremental protostellar activity drives the evaporation and sputtering of grain mantles, injecting sulfur into the gas phase and leading to a globally increasing trend in total gaseous sulfur abundance. Concurrently, the chemical evolution undergoes a distinct shift: while early stages are characterised by severe sulfur depletion, the more evolved phases exhibit a significant enrichment of oxygen-bearing species—such SO and SO$_2$—over carbon-bearing molecules, likely driven by an increased availability of atomic oxygen from water photodissociation. 

Investigating sulfur chemistry toward molecular cores using single-dish observations remains highly valuable (e.g., \citealt{vastel18,esplugues22,hily22,fernandez26,scholler26}), as it efficiently provides crucial large-scale statistics and serves as a foundation for advancing our understanding of molecular-species behaviour. However, interferometric observations are fundamental to unlocking the complexities of the chemistry in these small spatial-scale regions. Interferometric observations provide the high angular resolution needed to resolve individual cores, directly exposing the small-scale chemical pathways, ice-mantle sublimation zones, and chemical processes that usually remain averaged out within a single-dish beam.

In \citet{martinez24} and \citet{paron25}, using interferometric data from the Atacama Large Millimeter Array (ALMA), we investigated the sulfur-bearing species $^{34}$SO, SO, SO$^{+}$, SO$_{2}$, NS, and H$_{2}$CS toward a sample of 37 infrared-quiet ATLASGAL sources, representing the very early stages of star formation. Our findings show that the abundances of these molecules increase with gas temperature in the range 20--100 K, suggesting either progressive release from dust grains or activation of specific chemical pathways as temperature rises. Moreover, the strong correlations found among molecular abundances in those quiescent environments indicated a common chemical origin in the earliest stages of core evolution. Building upon these results, the present work aims to extend this systematic analysis to a sample of more evolved molecular cores. By comparing both results, we seek to characterise how sulfur chemistry increases in complexity (or not) as star formation progresses and the physical conditions of the cores become more dynamic.




Therefore, we present an analysis of a sample of 16 molecular cores in more advanced stages of star formation than our former studies to investigate the behaviour of the same sulfur-bearing species. In what follows, Sect.\,\ref{sectdata} presents the sample of analysed cores and the data used from the ALMA database. The results obtained from this analysis and the comparison with our previous findings are presented in Sect.\,\ref{sectresults}. In Sect.\,\ref{sectdiscuss}, we discuss the results and the implications for sulfur chemistry. This comparative approach lets us evaluate how the transition from quiescent, infrared-quiet stages to more active phases of star formation impacts the chemical evolution and gas-phase distribution of sulfur in the ISM.

\begin{table*}[h]
\centering
\caption{Sources sample and selected cores.}
\label{sample}
\begin{tabular}{lllcccc}
\hline
\hline
\#   & Source  &   v$_{\rm LSR}$ & \multicolumn{2}{c}{Core coordinates} &  $\theta_{\rm beam}$  & Dist. \\  
    &          & (km s$^{-1}$) & $\alpha$ (J2000) & $\delta$ (J2000)  &   (arcsec)              & (kpc)      \\
\hline
1  & G49.41+0.32   & -23.0  & 19:20:59.8 & 14:46:49.3  & 4.2 & 12.7\\
2  & G49.48-0.40   & +54.0   & 19:23:46.2 & 14:29:47.8  & 4.1 & 5.5 \\
3  & G213.71-12.60   & +9.5    & 06:07:47.7 & -06:22:56.3 & 3.8 & 0.8 \\
4  & G232.62+0.99  & +18.5   & 07:32:09.8 & -16:58:12.8 & 3.9 & 1.0\\
5  & G291.27-0.71    & -24.0  & 11:11:53.1 & -61:18:23.6 & 3.8 & 0.9 \\
6  & G294.52-1.62    & -15.0  & 11:35:32.2 & -63:14:43.7 & 3.8 & 1.4 \\
7  & G298.26+0.74    & -29.0  & 12:11:47.7 & -61:46:20.8 & 3.9 & 4.0 \\
8  & G305.21+0.21    & -40.3  & 13:11:13.7 & -62:34:40.9 & 3.8 & 3.1  \\
9  & G308.92+0.12    & -52.0  & 13:43:01.7 & -62:08:51.0   & 3.9 & 4.0 \\
10 & G309.92+0.48    & -59.7  & 13:50:41.8 & -61:35:10.7 & 3.9 &  5.2 \\
11 & G351.42+0.64    & -8.0   & 17:20:53.3 & -35:46:58.2 & 3.6 &  1.3\\
12 & G351.58-0.35    & -95.0  & 17:25:25.2 & -36:12:44.1 & 3.7 &  8.0\\
13 & G351.78-0.54    & -4.0   & 17:26:42.4 & -36:09:17.8 & 3.6 & 1.3\\
14 & G353.27+0.64  & -5.0   & 17:26:01.5 & -34:15:14.8 & 3.7 & 1.3\\
15 & G354.51+0.47    & -20.5  & 17:30:17.1 & -33:13:54.4 & 3.6 & 4.0\\
16 & G359.61-0.24    & +19.0   & 17:45:39.1 & -29:23:30.3 & 3.6 & 8.2\\
\hline
\end{tabular}
\end{table*}

\section{Source sample and data}
\label{sectdata}

From the ALMA database, we selected Project 2022.1.00974.S (PI:  Liu, Sheng-Yuan). This project comprises observations at Band 7 toward massive star-forming regions (MSFRs) associated with methanol masers. The presence of such masers ensures that cores embedded in these regions should be in a more advanced evolutionary stage of star formation than those analysed in our previous works (e.g., \citealt{elli06}). These data correspond to the same spectral windows as our previous study \citep{martinez24}, they are: spw15 (333.542--335.541 GHz), spw17 (335.481--337.48 GHz), spw19 (345.522--347.521 GHz), and spw21 (347.460--349.459 GHz).

Of all the observed regions in the project, we analysed those with a QA2 Status PASS, meaning the data passed ALMA Quality Assurance Level 2, ensuring reliable calibration for science-ready data. From that selection, from an inspection of the spectra toward the center of the cores, we then retained the ones exhibiting at least two of the following methanol spectral lines: 2(2,1)-3(1,2)(-\,-) at 335.133 GHz with $E_{\rm u}$=44.6 K, 12(1,11)-12(0,12)(-+) at 336.865 GHz with $E_{\rm u}$=197.07 K, and 14(7,8)-15(6,9)(++) at 336.438 GHz with $E_{\rm u}$=488.21 K. This allows us to calculate temperatures using the rotational diagram method (see Appendix\,\ref{appmetanol}) to thermally characterise each core. Cores were identified based on their continuum emission by inspecting the continuum maps at 0.87 mm. 
The source names (from the database), systemic velocities (v$_{\rm LSR}$), coordinates of the selected analysed molecular cores, beam size of the observation ($\theta_{\rm beam}$) for each case, and the distance are presented in Table\,\ref{sample}. The systemic velocities (v$_{\rm LSR}$) were obtained from the most intense and ubiquitous sulfur-bearing molecular line among the selected molecules, thioformaldehyde (H$_{2}$CS). Distances were compiled from an exhaustive search across the literature and catalogues in VizieR. For cores 6, 8, 10, 11, 12, and 13, distances were obtained from the ALMA QUARKS Survey catalogue \citep{liu24}. For the remaining sources, distances were retrieved from various methanol maser catalogues; where discrepancies arose, we cross-checked the indicated distances against distances derived from the Galactic rotation models of \citet{reid14} using the systemic velocities presented in Col.\,3.

\begin{table}[h]
\centering
\tiny
\caption{Analysed sulfur-bearing molecular lines.}
\label{lines}
\begin{tabular}{llcccc}
\hline
\hline
Molecule & Line & $\nu_{\rm rest}$ &   $E_{\rm u}$ & $g_{\rm u}$ & log($A_{\rm ul}$)  \\
         &      &   (GHz)           &     (K)      &         &  (s$^{-1}$) \\  
\hline
 & &  & & & \\[-1.8ex]
$^{34}$SO                   & 7(8)--6(7)            & 333.900  & 79.8  & 15  & -3.32     \\
SO$_{2}$ v=0                & 8(2,6)--7(1,7)        & 334.673  & 43.1  & 17  & -3.89   \\
NS v=0                      & 15/2--13/2            & 346.221  & 69.8  & 16  & -3.14  \\
SO v=0 $^{3}$$\Sigma$       & 9(8)--8(7)            & 346.528  & 78.7  & 19  & -3.26   \\
SO$^{+}$                    & 15/2--13/2 (1/2) l=f  & 348.115  & 70.2  & 16  & -3.64  \\
H$_{2}$CS                   &  10(1,9)--9(1,8)      & 348.534  & 105.2 & 21  & -3.20   \\
\hline

 & &  &   & &    \\[-1.8ex]

\end{tabular}
\end{table}

The continuum maps and data cubes correspond to observations performed in the 7\,m array. 
The observed frequency range and the spectral resolution are 333.3–349.1 GHz and 1.1 MHz, respectively. The angular resolution goes from 3$\farcs$6 to 4$\farcs$2  for the selected sources (see Col.\,6 in Table\,\ref{sample}). The sensitivities are between 0.5 and 1.7 mJy beam$^{-1}$ for the continuum, and between 19 and 43 mJy beam$^{-1}$ for the line emission (each 10 km\,s$^{-1}$).

We analysed the same sulfur-bearing molecular lines as in our previous work. Table\,\ref{lines} presents the molecules, transitions, rest frequencies, degeneracy of the upper level, and the Einstein coefficient obtained from the Splatalogue Database and the Cologne Database for Molecular Spectroscopy (CDMS; \citealt{muller05}).

\begin{table*}[h!]
\centering
\small
\caption{Column densities ($\times 10^{14}$ cm$^{-2}$).}
\label{coldens}
\begin{tabular}{lcccccccccccc}
\hline
\hline
\# & $N$($^{34}$SO) & err. & $N$(NS) & err. & $N$(SO$^+$) & err. & $N$(H$_2$CS) & err. & $N$(SO$_2$) & err. & $N$(SO) & err. \\
\hline
1  & 0.11 & 0.01 & -     & -     & -     & -     & 1.45   & 0.08 & 2.95   & 0.23  & 0.95  & 0.07 \\
2  & 0.77 & 0.10 & 0.74  & 0.09  & 0.31  & 0.04  & 7.35   & 0.43 & 5.20   & 0.48  & 5.14  & 0.27 \\
3  & 1.56 & 0.11 & -     & -     & 0.42  & 0.04  & 0.78   & 0.04 & 26.00  & 1.82  & 7.52  & 0.86 \\
4  & 0.05 & 0.01 & 0.13  & 0.01  & -     & -     & 1.74   & 0.19 & 0.59   & 0.08  & 0.91  & 0.05 \\
5  & 1.90 & 0.25 & -     & -     & 1.18  & 0.14  & 3.02   & 0.25 & 29.00  & 1.90  & 6.65  & 0.97 \\
6  & 0.35 & 0.03 & -     & -     & -     & -     & 2.09   & 0.12 & 4.91   & 0.29  & 3.03  & 0.41 \\
7  & 1.15 & 0.13 & -     & -     & 0.41  & 0.05  & 1.43   & 0.18 & 31.40  & 3.25  & 4.39  & 0.65 \\
8  & 4.32 & 0.38 & 11.20 & 1.18  & -     & -     & 43.80  & 5.82 & 131.00 & 14.65 & 23.30 & 3.17 \\
9  & 1.43 & 0.15 & 1.21  & 0.15  & 0.50  & 0.03  & 5.38   & 0.39 & 30.60  & 2.42  & 6.50  & 0.38 \\
10 & 0.70 & 0.05 & 1.84  & 0.11  & 0.32  & 0.02  & 13.70  & 1.56 & 17.30  & 1.50  & 4.34  & 0.38 \\
11 & 30.50& 2.16 & 74.40 & 5.71  & 22.10 & 3.18  & 251.00 & 28.82& 621.00 & 68.88 & 59.70 & 4.01 \\
12 & 9.97 & 1.23 & 31.30 & 2.08  & 4.65  & 0.41  & 93.20  & 13.88& 224.00 & 25.54 & 39.80 & 4.21 \\
13 & 37.10& 4.39 & 122.00& 16.38 & 13.50 & 1.72  & 156.00 & 11.37& 783.00 & 41.66 & 97.10 & 7.92 \\
14 & 0.41 & 0.03 & 2.78  & 0.20  & 0.92  & 0.13  & 14.40  & 1.98 & 6.34   & 0.52  & 2.66  & 0.31 \\
15 & 1.15 & 0.10 & 5.23  & 0.74  & -     & -     & 23.60  & 2.26 & 29.00  & 2.22  & 8.03  & 0.60 \\
16 & 0.89 & 0.09 & 3.66  & 0.28  & -     & -     & 17.60  & 1.91 & 25.40  & 3.55  & 7.84  & 0.71 \\
\hline
\multicolumn{13}{l}{A `-' means that the emission of the molecule was not observed, hence no column density was derived.} \\
\end{tabular}
\end{table*}

\section{Results}
\label{sectresults}

We extracted spectra from a beam-size region centred at the continuum peak position of each core. As an illustrative example, the case for the region G351.58-0.35 (core 12) is presented in Appendix\,\ref{example}. In all cases, the peaks of the molecular emission match those of the continuum emission. Gaussian fittings to each line were applied to obtain the intensity peaks, the full width at half maximum (FWHM) $\Delta$v, and the integrated line intensity (i.e., the line flux, $W$).  All the parameters obtained from the Gaussian fittings are presented in Tables\,\ref{gauss1}, \ref{gauss2}, and \ref{gauss3} in Appendix\,\ref{appendGauss}.   

To estimate the column densities of each molecular species, assuming local thermodynamic equilibrium (LTE), following \citet{artur23}, we first derived the column densities of the
upper levels, N$_{\rm u}$ (in cm$^{-2}$), of each transition from

\begin{equation}
{N_{\rm u} = 2375 \times 10^6~\left(\frac{W}{\rm Jy~beam^{-1}~km~s^{-1}}\right)~\left(\frac{\rm s^{-1}}{A_{\rm ul}}\right)~\left(\frac{\rm arcsec}{\theta_{\rm beam}}\right)^{2}}
\label{coldensEq}
,\end{equation}

\noindent where $W$ is the measured line flux in Jy beam$^{-1}$ km s$^{-1}$, $A_{\rm ul}$ is the line Einstein coefficient (in s$^{-1}$) (see Table \ref{lines}), and  $\theta_{\rm beam}$ is the beam of the observations (Col\,6 in Table\,\ref{sample}).

Then, the total column density of each molecule was obtained from

\begin{equation}
N = \frac{N_{\rm u}~Q(T_{\rm ex})~\exp(E_{\rm u}/T_{\rm ex})}{g_{\rm u}}
\label{coldensEq}
,\end{equation}

\noindent where $Q(T_{\rm ex})$ is the rotational partition function, $T_{\rm ex}$ is the excitation temperature (in K), $g_{\rm u}$ the upper-level degeneracy, and $E_{\rm u}$ (in K) the upper energy of the line transition.  All these parameters, except $Q$, are shown in Table\,\ref{lines}. 
$T_{\rm ex}$ was assumed to be equal to the kinetic temperature ($ T_{\rm k}$) of the gas in each core, which was assumed to be equal to the $T_{\rm rot}$ derived from the methanol (see Appendix\,\ref{metanol}). The temperature-dependent partition function was extrapolated to the measured temperature using the information provided in the CDMS catalogue via the Splatalogue database.
All molecular column densities are presented in Table\,\ref{coldens}.


To evaluate the possible impact of line opacity on the derived column densities, we estimated the optical depth ($\tau$) for the SO and SO$_2$ by following the procedure described in \cite{martinez24}. As noted by previous studies of sulfur chemistry in shocked hot cores \citep[e.g.,][]{esplugues2013}, optical depth effects can lead to underestimating column densities if uncorrected. However, our calculations show that the transition SO v=0 $^{3}$$\Sigma (9_8-8_7)$ is optically thin ($\tau < 0.3$) across all 16 sources. Similarly, for 13 of the 16 cores, the $\text{SO}_2$ ($8_{2,6}-7_{1,7}$) transition also remains optically thin ($\tau < 0.02$), yielding an opacity correction factor $C_{\tau} = \tau / (1 - e^{-\tau}) \approx 1$. Moderate opacity ($\tau \sim 0.8 - 2.0$) is only retrieved for $\text{SO}_2$ in cores G351.42+0.64 (\#11), G351.58$-$0.35 (\#12), and G351.78$-$0.54 (\#13). Incorporating $C_{\tau}$ in these three cases leads to a modest increase in their SO$_2$ column densities by factors of 1.5 to 2.3, an effect that remains well within the uncertainties and does not impact the overall statistical trends or abundance ratios discussed below.

We measure the flux density ($S_{\rm \nu}$) from a beam toward the peak of each core from the 0.8 mm continuum emission maps (Col.\,2 in Table\,\ref{coldensh2}).  Using these values we
derive the H$_{2}$ column density, $N$(H$_{2}$), of each core. To carry on this, we use the following equation \citep{kauff08}:

\begin{eqnarray}
N({\rm H_{2}})=2.02 \times 10^{20}~{\rm cm^{-2}} \left(\exp^{1.439(\lambda/{\rm mm})^{-1}(T_{\rm dust}/10~{\rm K})^{-1}} - 1 \right) \nonumber \\ 
 \times \left( \frac{\lambda}{mm}\right)^{3}\left(\frac{\kappa_{\nu}}{0.01 ~{\rm cm^{2}g^{-1}}} \right)^{-1}\left(\frac{S_{\nu} }{\rm mJy~beam^{-1}} \right) \left(  \frac{\theta_{\rm beam}}{10~{\rm arcsec}} \right)^{-2}    
\label{NH2}
\end{eqnarray}

\noindent where $T_{\rm dust}$ is the dust temperature and $\kappa_{\nu}$ is the dust opacity per gram of matter at 870~$\mu$m, for which we adopt the value of 0.0185~cm$^2$g$^{-1}$ \citep[][and references therein]{csengeri2017}. We assume thermal coupling between dust and gas (i.e., $T_{\rm dust}=T_{\rm kin}$). The obtained $N$(H$_{2}$) values are included in Table\,\ref{coldensh2}. Finally, we obtain the abundance of each molecular species, $X$=$N$(species)/$N$(H$_2$), and perform a statistical analysis to investigate potential correlations among distinct molecular abundances, as well as between abundances and temperature of each core. Table\,\ref{abund} presents the obtained logarithmic molecular abundances.

\begin{table}[h]
\centering
\tiny
\caption{Measured flux densities and derived H$_{2}$ column densities. }
\label{coldensh2}
\begin{tabular}{lcccccccc}
\hline
\hline
\# & $S_{\rm \nu}$ & err. & $N$(H$_2$) & err. \\
 & (Jy beam$^{-1}$) &  & ($\times 10^{22}$ cm$^{-2}$) &  & \\
 \hline
1  & 0.27  & 0.02 & 0.73  & 0.09 & \\
2  & 2.21  & 0.22 & 36.27 & 5.46 & \\
3  & 0.29  & 0.03 & 1.52  & 0.22 & \\
4  & 0.35  & 0.03 & 6.31  & 0.89 & \\
5  & 1.39  & 0.08 & 7.34  & 0.87 & \\
6  & 0.51  & 0.03 & 8.25  & 1.04 & \\
7  & 0.39  & 0.02 & 1.60  & 0.18 & \\
8  & 2.88  & 0.27 & 11.95 & 1.67 & \\
9  & 0.25  & 0.02 & 1.00  & 0.13 & \\
10 & 1.15  & 0.10 & 4.70  & 0.63 & \\
11 & 21.96 & 1.12 & 69.38 & 7.92 & \\
12 & 5.03  & 0.50 & 21.77 & 3.18 & \\
13 & 10.55 & 0.97 & 51.23 & 7.09 & \\
14 & 2.27  & 0.14 & 7.49  & 0.90 & \\
15 & 2.89  & 0.17 & 14.24 & 1.69 & \\
16 & 1.63  & 0.10 & 9.21  & 1.18 & \\
\hline
\end{tabular}
\end{table}


\begin{table*}[h]
\centering
\small
\caption{Logarithmic molecular abundances and propagated error.}
\label{abund}
\begin{tabular}{lcccccccccccc}
\hline
\hline
\# & $X$($^{34}$SO) & err. & $X$(NS) & err. & $X$(SO$^+$) & err. & $X$(H$_2$CS) & err. & $X$(SO$_2$) & err. & $X$(SO) & err. \\
\hline
1  & -8.82 & 0.07 & -     & -     & -      & -     & -7.70 & 0.06 & -7.39 & 0.06 & -7.89 & 0.06 \\
2  & -9.67 & 0.09 & -9.69 & 0.08 & -10.07 & 0.09 & -8.69 & 0.07 & -8.84 & 0.08 & -8.85 & 0.07 \\
3  & -7.99 & 0.07 & -     & -     & -8.56  & 0.08 & -8.29 & 0.07 & -6.77 & 0.07 & -7.31 & 0.08 \\
4  & -10.10& 0.11 & -9.69 & 0.07 & -      & -     & -8.56 & 0.08 & -9.03 & 0.08 & -8.84 & 0.07 \\
5  & -8.59 & 0.08 & -     & -     & -8.79  & 0.07 & -8.39 & 0.06 & -7.40 & 0.06 & -8.04 & 0.08 \\
6  & -9.37 & 0.07 & -     & -     & -      & -     & -8.60 & 0.06 & -8.23 & 0.06 & -8.44 & 0.08 \\
7  & -8.14 & 0.07 & -     & -     & -8.59  & 0.07 & -8.05 & 0.07 & -6.71 & 0.07 & -7.56 & 0.08 \\
8  & -8.44 & 0.07 & -8.03 & 0.08 & -      & -     & -7.44 & 0.08 & -6.96 & 0.08 & -7.71 & 0.08 \\
9  & -7.84 & 0.07 & -7.92 & 0.08 & -8.30  & 0.06 & -7.27 & 0.06 & -6.51 & 0.07 & -7.19 & 0.06 \\
10 & -8.83 & 0.07 & -8.41 & 0.06 & -9.17  & 0.06 & -7.54 & 0.08 & -7.43 & 0.07 & -8.03 & 0.07 \\
11 & -8.36 & 0.06 & -7.97 & 0.06 & -8.50  & 0.08 & -7.44 & 0.07 & -7.05 & 0.07 & -8.07 & 0.06 \\
12 & -8.34 & 0.08 & -7.84 & 0.07 & -8.67  & 0.07 & -7.37 & 0.09 & -6.99 & 0.08 & -7.74 & 0.08 \\
13 & -8.14 & 0.08 & -7.62 & 0.08 & -8.58  & 0.08 & -7.52 & 0.07 & -6.82 & 0.06 & -7.72 & 0.07 \\
14 & -9.26 & 0.06 & -8.43 & 0.06 & -8.91  & 0.08 & -7.72 & 0.08 & -8.07 & 0.06 & -8.45 & 0.07 \\
15 & -9.09 & 0.06 & -8.44 & 0.08 & -      & -     & -7.78 & 0.07 & -7.69 & 0.06 & -8.25 & 0.06 \\
16 & -9.01 & 0.07 & -8.40 & 0.06 & -      & -     & -7.72 & 0.07 & -7.56 & 0.08 & -8.07 & 0.07 \\
\hline
\multicolumn{13}{l}{A `-' means that \textbf{ the emission of the molecule was not observed}, hence no $X$ was derived.}\\
\end{tabular}
\end{table*}

Following previous approaches such as \citet{rodriguez-b21}, we employed a Spearman rank correlation matrix to examine potential monotonic associations between the abundances and kinetic temperatures. Utilising this non-parametric Spearman coefficient ($\rho$) allows us to assess possible monotonic trends without assuming just some kind of linear correlation. The spearman coefficient ranges from $-1$ to $+1$ and describes the direction and strength of the association between two variables: $\rho=$$+1$ indicates a perfect positive monotonic relationship, $\rho=$$-1$ indicates a perfect negative monotonic relationship, and $\rho$$=0$ indicates no monotonic association. Thus, values of $|\rho|$ closer to 1 indicate stronger associations, whereas values closer to 0 indicate weaker associations. To ensure the reliability of the correlations, to test the null hypothesis we calculated $p$-values to assess whether the observed associations could be explained by chance. Low $p$-values provide strong evidence that the observed association is unlikely to occur in the absence of a real association.

\begin{figure}[h]
    \centering
    \includegraphics[width=1\linewidth]{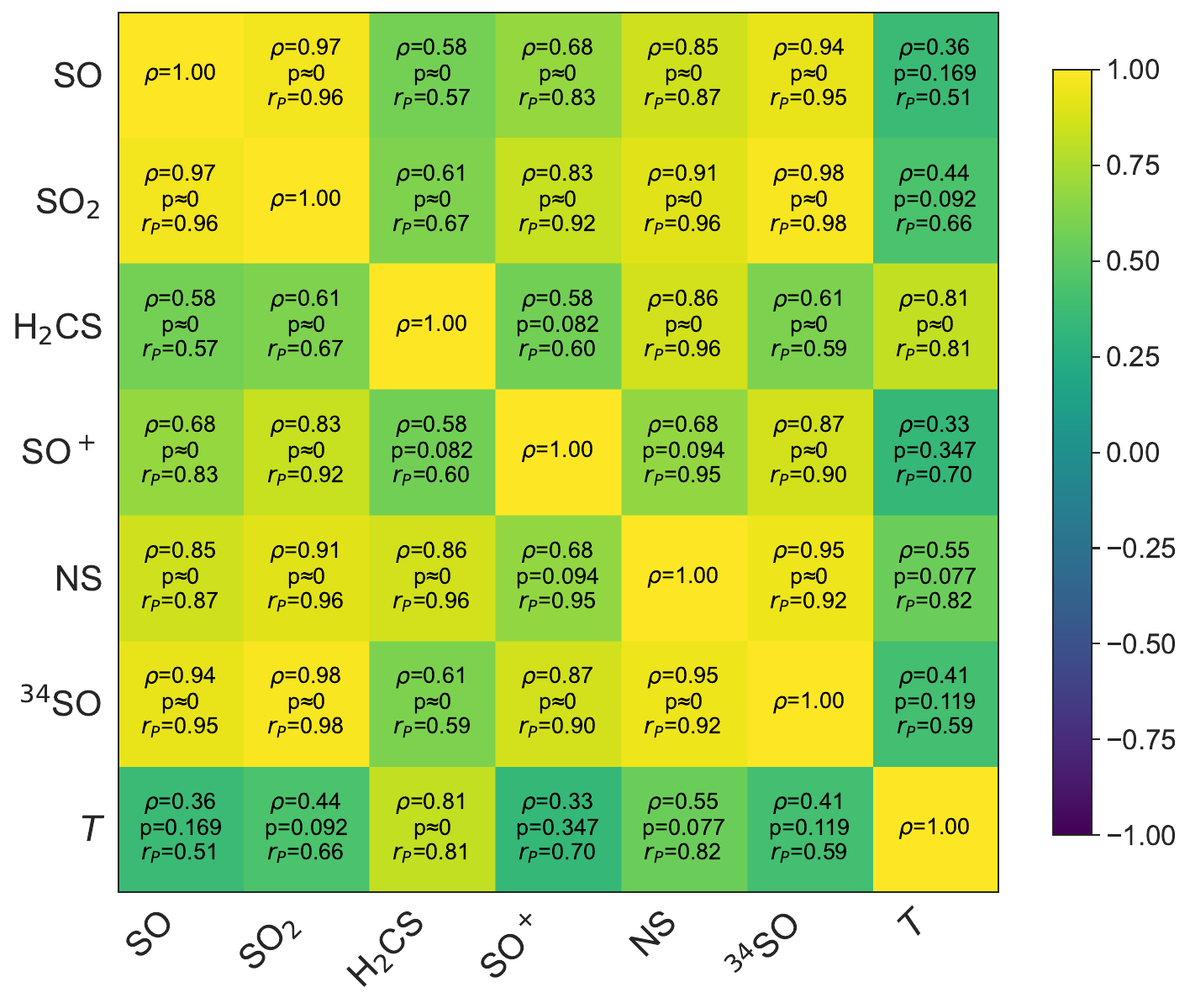}
    \caption{Correlation matrix showing the relationships among molecular abundances and between abundances and temperature. The colour scale indicates the Spearman coefficient, which is also indicated in each box. $p$-values of such correlations are included ($p$$\approx$0 means that the values are lower than 0.001). Additionally, Pearson coefficients ($r_{p}$) are also reported for each case.} 
    \label{spearman}
\end{figure}

\begin{figure}[h]
    \centering
    \includegraphics[width=1\linewidth]{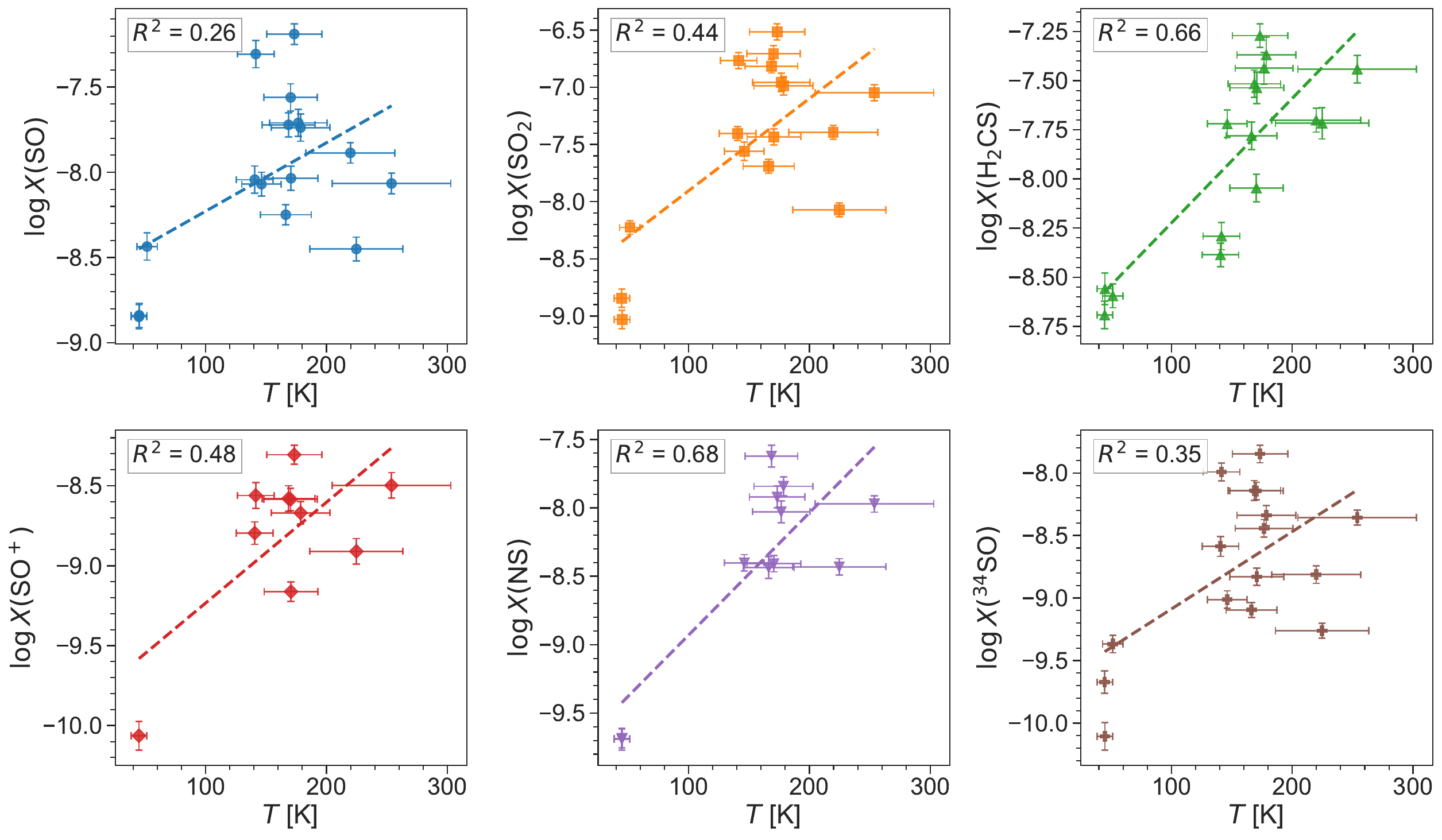}
    \caption{Molecular abundances vs kinetic temperature for the analysed core sample. Dashed lines indicate linear fits. The corresponding \(R^2\) values are included in each panel.}
    \label{scatterSulfurT}
\end{figure}

Figure\,\ref{spearman} presents the Spearman correlation matrix for the abundances and temperature across the core sample, along with their corresponding $p$-values and Pearson coefficients to assess linearity. To complement this statistical analysis, Fig.\,\ref{scatterSulfurT} shows the individual abundances versus kinetic temperature with linear fits and their associated $R^2$ values. While the Spearman coefficient evaluates the strength and direction of monotonic trends regardless of linearity, $R^2$ quantify how effectively a straight line describes the data. Together with the Pearson coefficient, it serves as an indicator of the degree to which the observed chemical trends adhere to a linear relationship.

\begin{figure}[h]
    \centering
    \includegraphics[width=1\linewidth]{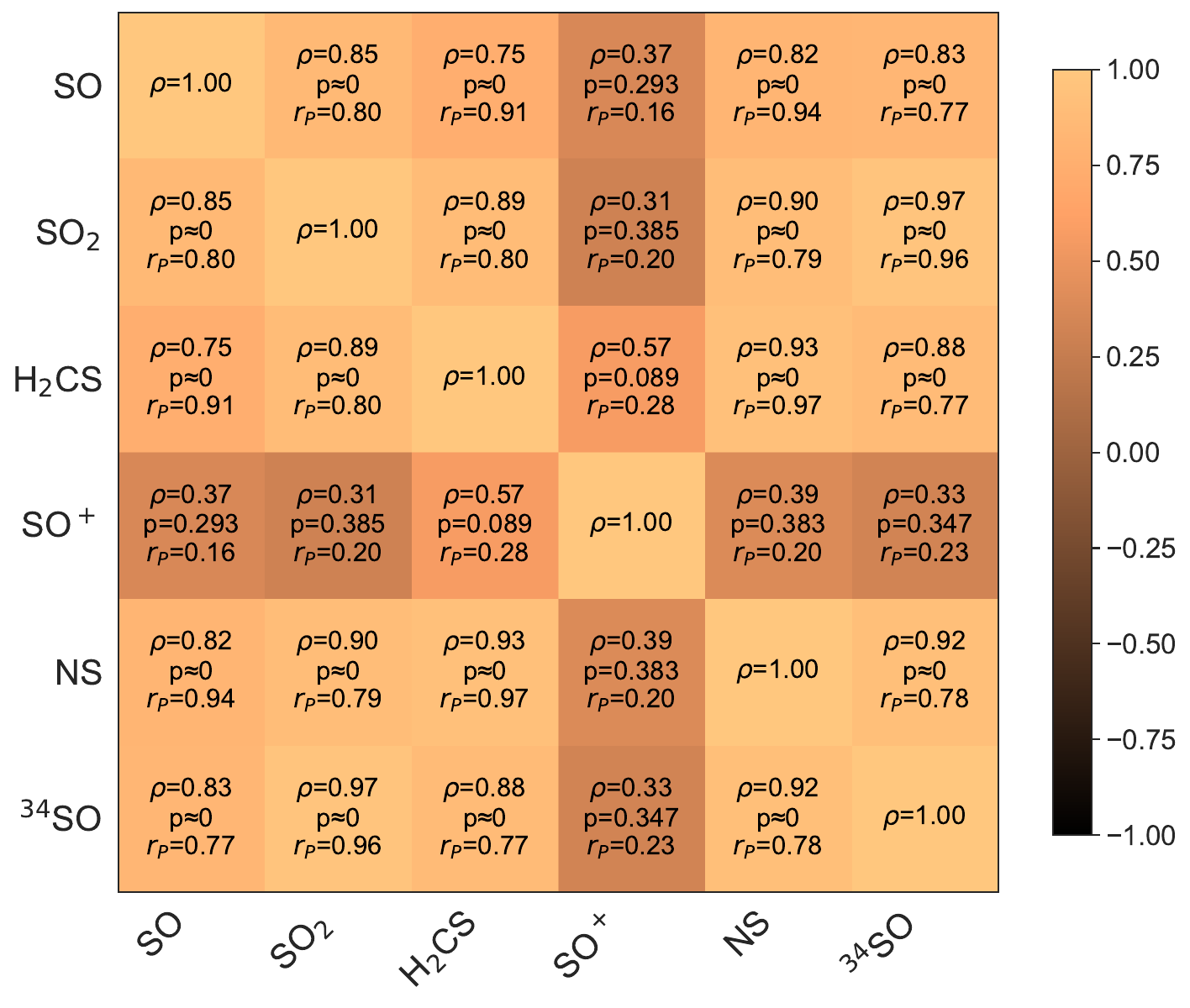}
    \caption{Correlation matrix showing the relationships between the line widths ($\Delta$v) measured for the different sulfur-bearing molecular lines. The colour scale indicates the Spearman coefficient, which is also indicated in each box. $p$-values of such correlations are included ($p$$\approx$0 means that the values are lower than 0.001). Additionally, Pearson coefficients ($r_{p}$) are also reported for each case.}
    \label{spearmanDelta}
\end{figure}

In addition, we studied correlations between line widths ($\Delta$v) of the analysed sulfur-bearing species. Figure\,\ref{deltaVfig} presents the comparisons of the line widths in typical scatter plots, and Fig.\,\ref{spearmanDelta} displays the Spearman correlation matrix between $\Delta$v of each molecular line. As done in Fig.\,\ref{spearman}, the $p$-values of the Spearman correlations and Pearson coefficients ($r_{p}$) are also included.

\subsection{Comparisons between early and evolved cores}

In what follows, we present results from the comparisons between two samples of cores at different evolutionary stages. We compare the results from the cores presented here (evolved sample) with results presented in \citet{martinez24} (early sample).

\subsubsection{Abundances}

\begin{figure}[h]
    \centering
    \includegraphics[width=1\linewidth]{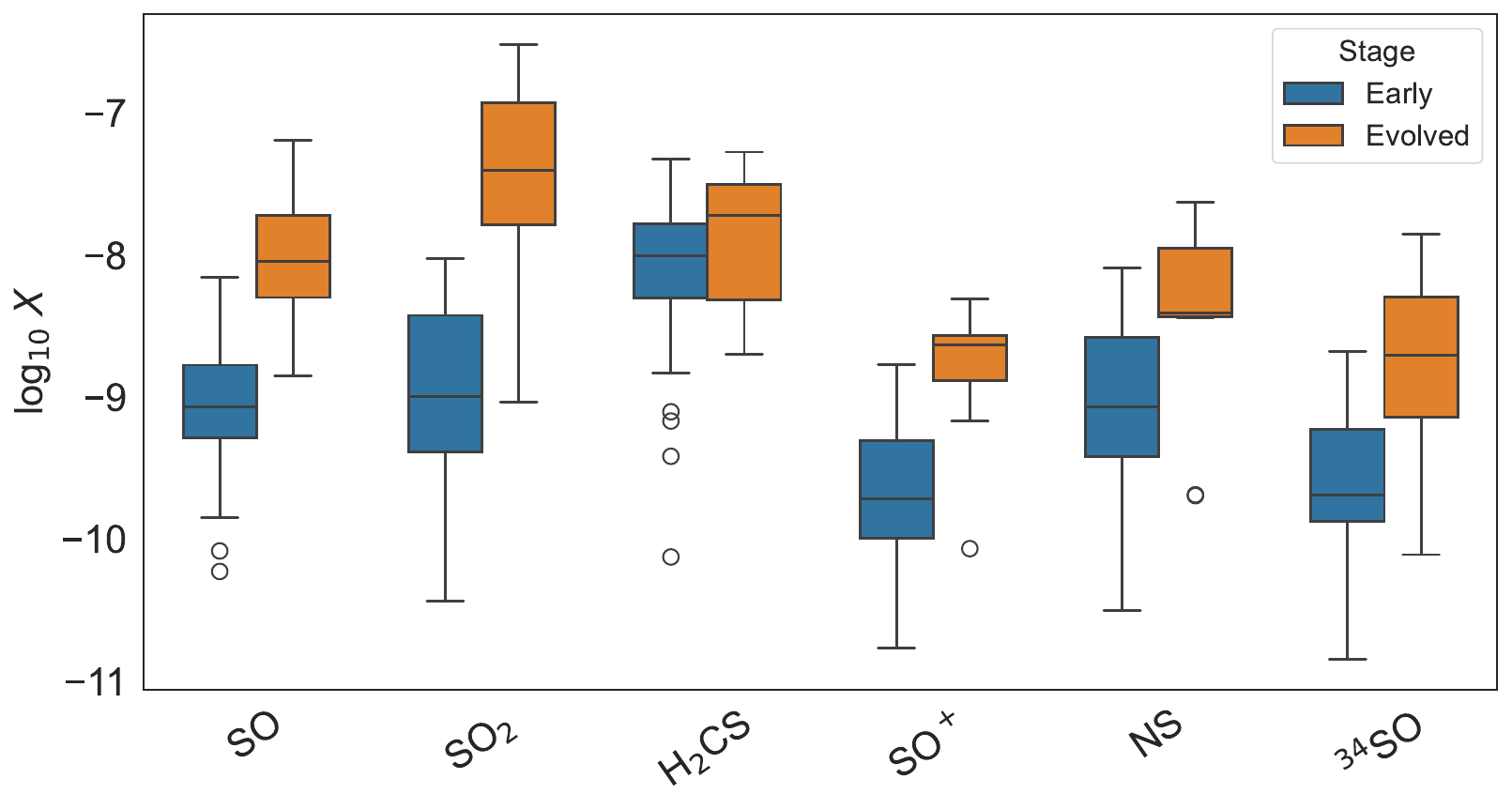}
   \caption{Comparison of sulfur-bearing molecular abundances between early and evolved cores. For each molecule, the horizontal line inside the box indicates the median log($X$), the box spans the interquartile range, and the whiskers show the range of non-outlier values. Outliers are shown as circles beyond the whiskers.}
    \label{fig:abundancias_sulfuradas}
\end{figure}

We computed the median abundances of sulfur-bearing species for the sample of evolved and early cores. This is shown in Fig.\,\ref{fig:abundancias_sulfuradas}. The corresponding values are listed in Table~\ref{tab:sulphur_abundances_early_evolved}, where $N_{\rm early}$ and $N_{\rm evolv.}$ indicate the number of valid abundance measurements in the early and evolved samples, respectively. The columns Med\,X$_{\rm early}$ and Med\,X$_{\rm evolv.}$ give the median abundances in each group, while $f_{\rm evolv./early}$ is the ratio between the evolved and early median values. The last column gives the Mann--Whitney $p$-value, used to evaluate whether the abundance distributions of two samples are statistically different.

\begin{table}[h]
\centering
\caption{Comparison of sulfur-bearing molecular abundances between early and evolved cores.}
\label{tab:sulphur_abundances_early_evolved}
\small
\setlength{\tabcolsep}{3pt}
\begin{tabular}{lcccccc}
\hline \hline
Mol. & $N_{\rm early}$ & $N_{\rm evolv.}$ & 
Med\,$X_{\rm early}$ & Med\,$X_{\rm evolv.}$ & 
$f_{\rm evolv./early}$ & $p_{\rm MW}$ \\
\hline
SO$_2$  & 32 & 16 & $1.02{\times}10^{-9}$  & $4.00{\times}10^{-8}$ & 39.34 & $1.76{\times}10^{-6}$ \\
SO$^+$  & 25 & 10 & $1.94{\times}10^{-10}$ & $2.37{\times}10^{-9}$ & 12.22 & $3.71{\times}10^{-4}$ \\
SO      & 35 & 16 & $8.61{\times}10^{-10}$ & $9.15{\times}10^{-9}$ & 10.62 & $8.53{\times}10^{-7}$ \\
$^{34}$SO & 32 & 16 & $2.08{\times}10^{-10}$ & $2.07{\times}10^{-9}$ & 9.95  & $1.92{\times}10^{-4}$ \\
NS      & 34 & 11 & $8.67{\times}10^{-10}$ & $3.97{\times}10^{-9}$ & 4.58  & $2.49{\times}10^{-3}$ \\
H$_2$CS & 36 & 16 & $1.00{\times}10^{-8}$  & $1.92{\times}10^{-8}$ & 1.92  & $1.45{\times}10^{-1}$ \\
\hline
\end{tabular}
\end{table}

\begin{figure}[h]
    \centering

    \includegraphics[width=0.95\linewidth]{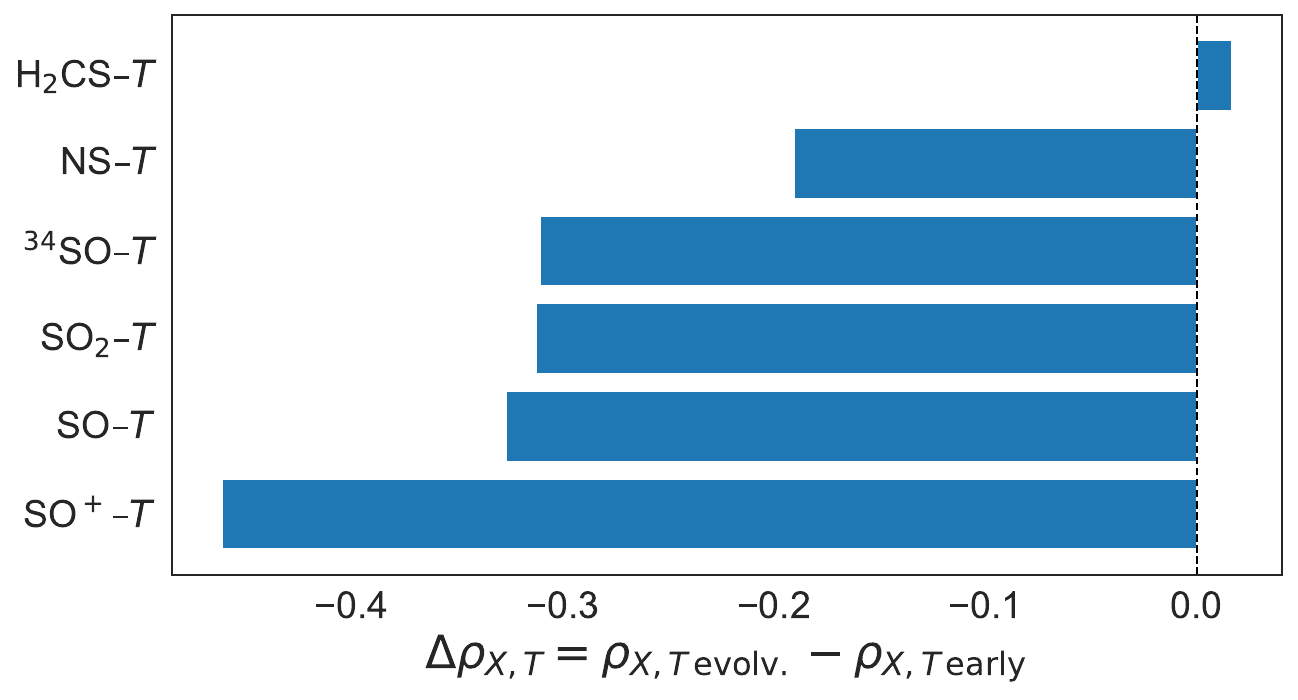}

    \vspace{0.3cm}

    \includegraphics[width=0.95\linewidth]{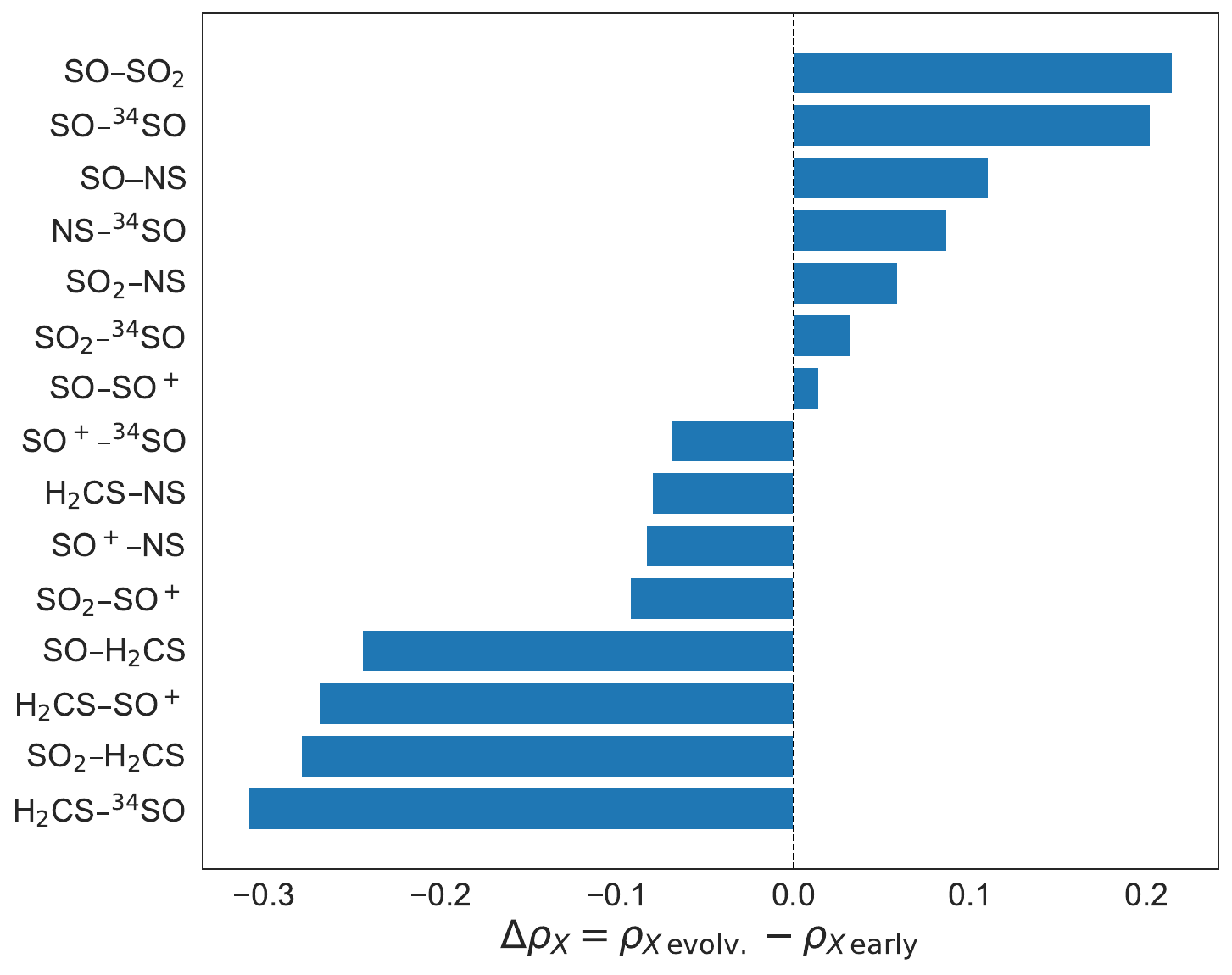}

    \caption{
    Difference between the Spearman correlation coefficients of the evolved and early samples.
    Top panel: abundance--temperature pairs, quantified as
    $\Delta \rho_{X,T} =
    \rho_{X,T\,\mathrm{evolv.}} -
    \rho_{X,T\,\mathrm{early}}$.
    Bottom panel: abundance--abundance pairs, quantified as
    $\Delta \rho_{X} =
    \rho_{X\,\mathrm{evolv.}} -
    \rho_{X\,\mathrm{early}}$.
    }
    \label{fig:diff_spearman_abundances}
\end{figure}

We compared the correlations obtained above for the analysed evolved cores with those corresponding to the early-stage ones. To do this, we constructed the Spearman correlation matrices between abundances and temperature using the results obtained for the early-core sample analysed by \citet{martinez24}. For the sake of space, these matrices are not included here, but it is important to remark that the obtained $p$-values are extremely close to zero. Then, we computed the difference between the Spearman correlation coefficients of the evolved and early samples, defined as $\Delta \rho_{\rm X,T}$ and $\Delta \rho_{\rm X}$ for the comparisons between abundance and temperature and among abundances (upper and lower panels of Fig.\,\ref{fig:diff_spearman_abundances}, respectively). Positive values of $\Delta \rho$ indicate that a given parameter pair is more strongly correlated in the evolved cores, while negative values indicate that the correlation is weaker in the evolved sample. 

\begin{figure*}[h]
    \centering
    \includegraphics[width=0.8\linewidth]{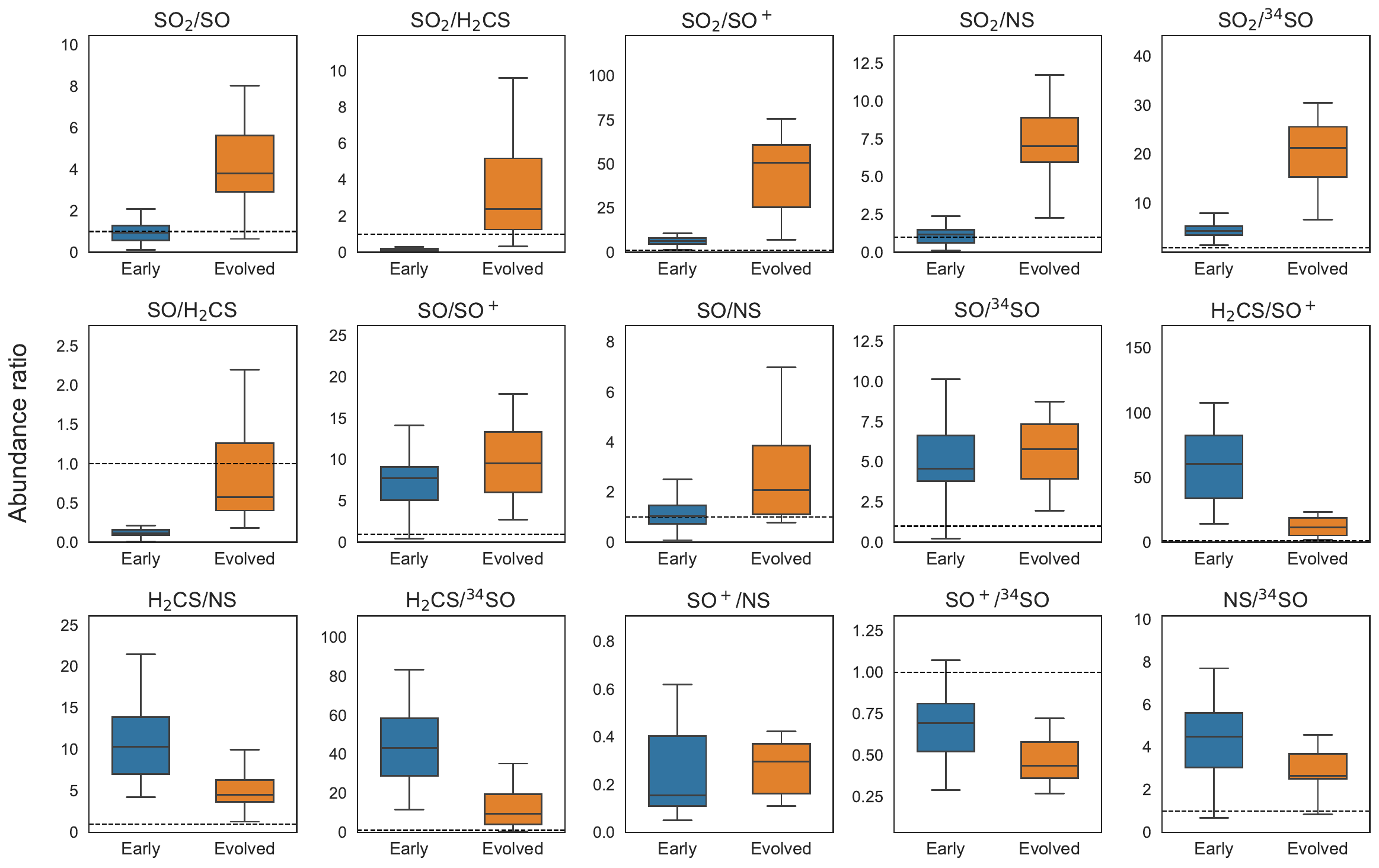}
    \caption {Comparison of the sulfur-bearing abundance ratios between early and evolved cores. 
    The horizontal
line inside the box indicates the median ratio, the box spans
the interquartile range, and the whiskers show the range of non-outlier values.
    The dashed horizontal line indicates a ratio equal to unity. Outliers are not displayed for visual clarity. Outliers are shown as circles beyond the whiskers.}
    \label{fig:ratios_sulfuradas}
\end{figure*}

Finally, we explored molecular abundance ratios to evaluate whether the observed abundance enhancement in evolved cores is proportional across all species or if specific molecules become dominant. All possible ratio comparisons are presented in Fig.\,\ref{fig:ratios_sulfuradas}. Those ratios involving SO and SO$_2$ are particularly noteworthy, as these species are widely recognised in the literature as reliable chemical clocks for tracking core evolution.

\subsubsection{Line widths}

\begin{figure}[h]
    \centering
    \includegraphics[width=1\linewidth]{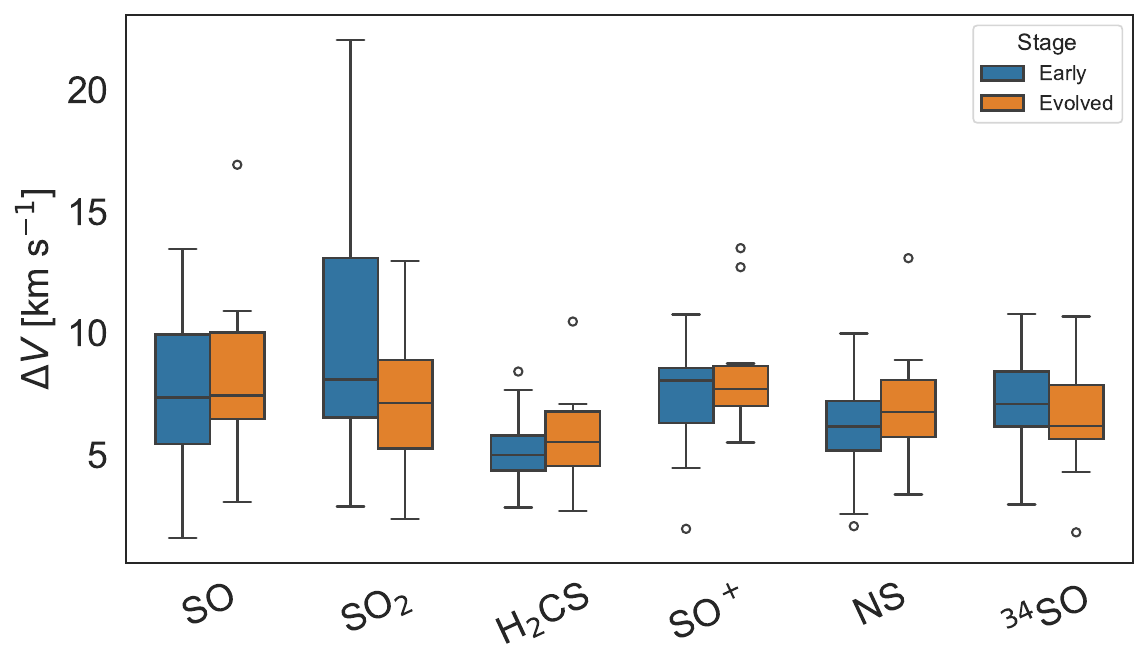}
    \caption{Comparison of line widths of the sulfur-bearing species lines between the early and evolved core samples. The molecules are ordered according to the spatial distribution proposed by \citet{martinez24}, from species expected to trace more central regions to species likely associated with more external layers. The horizontal
line inside the box indicates the median $\Delta$v, the box spans
the interquartile range, and the whiskers show the range of non-outlier values. Outliers are shown as circles beyond the whiskers.}
    \label{deltaV_medians}
\end{figure}

\begin{figure}[h]



        \centering
        \includegraphics[width=1\linewidth]{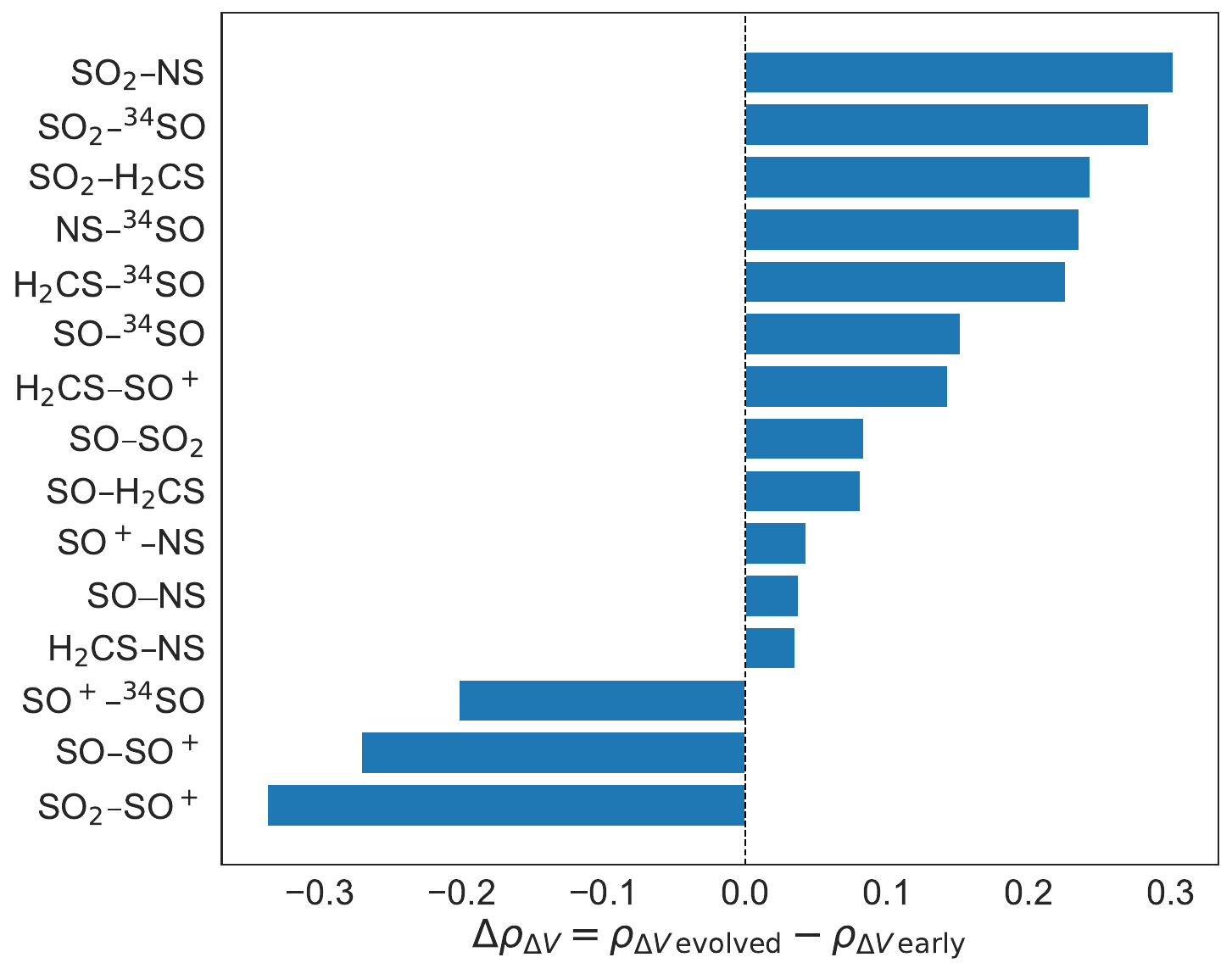}
        \label{fig:diff_spearman_deltav_pairs}

    \caption{Comparison of the changes in the Spearman correlation coefficients of the line widths between evolved and early cores. 
    The values correspond to $\Delta \rho_{\rm \Delta v} = \rho_{\rm \Delta v\,evolv.} - \rho_{\rm \Delta v\,early}$. Positive values indicate stronger correlations in the evolved sample, while negative values indicate weaker correlations.
}
    \label{diffdeltav}
\end{figure}

We also compared the line widths ($\Delta$v) of the sulfur-bearing molecular transitions between the early and evolved core samples. Figure\,\ref{deltaV_medians} shows the median $\Delta$v values for both groups. The molecules are arranged according to the spatial distribution proposed by \citet{martinez24}, from species expected to arise closer to the central regions of the cores to species likely associated with more external layers.
As with abundances, we also computed the difference between the Spearman correlation coefficients between the line widths ($\Delta \rho$$_{\Delta {\rm v}}$) of the different species obtained from the evolved and early core samples. Figure\,\ref{diffdeltav} presents the resulting $\Delta \rho$$_{\Delta {\rm v}}$.

\subsection{Modelling and chemical age}
\label{models}

To interpret the observed molecular abundances and evaluate the physical mechanisms controlling sulfur chemistry within the core sample, astrochemical simulations were performed using the Nautilus gas-grain code \citep{ruaud16}, and the 2024 KIDA network for interstellar astrochemistry \citep{wake24}. We implemented the same framework for physical conditions as \citet{tani19} in their chemical simulations of hot cores. A three-phase (gas, bulk, and mantle) evolutionary model was adopted, beginning with a cold collapse stage followed by a warm-up phase. The cold collapse phase lasts for $5 \times 10^{5}$~years at a constant temperature of 10~K. During this initial stage, the gas density ($ n_{\rm H}$) increases from a typical cloud value of $10^{4}$~cm$^{-3}$ to a dense regime of $10^{7}$~cm$^{-3}$. In tandem with this density increase, the visual extinction ($A_{\rm V}$) rises from 5~mag to a highly shielded value of 500~mag. Following the formulation of \citet{tani19}, the temperature during the warm-up phase rises from 10~K to a maximum of 200~K based on a quadratic heating scale with time. Incorporating this structural evolution in the simulation is critical to properly capture the physical transition from a cold, quiescent prestellar environment to the active, heated core of a massive protostar.
To evaluate how the timescale impacts the chemistry, three distinct warm-up scenarios were tested: a fast heating scale ($5 \times 10^{4}$~years), an intermediate scale ($2 \times 10^{5}$~years), and a slow scale ($1 \times 10^{6}$~years).
For the initial chemical conditions of the collapse, the following cosmic elemental abundances were adopted: $1.5 \times 10^{-5}$ for sulfur, $2.4 \times 10^{-4}$ for oxygen, and $1.7 \times 10^{-4}$ for carbon \citep{hincelin11,vidal17}.
Regarding the sulfur abundance, while a moderate depletion factor could be considered (typically between 10 and 20), we decided to run our simulations without it because the $5 \times 10^5$ year cold collapse phase included in our model already naturally drives the progressive freeze-out of gaseous sulfur onto dust grains as H$_{2}$S, HS, and S$_{8}$ \citep{wake24}. Therefore, imposing a depletion factor at $t=0$ would artificially duplicate this physical process. Moreover, \citet{fuente23} note that sulfur depletion is highly sensitive to the local environment; hence, to maintain consistency across comparisons among the analysed sources, we used the sulfur cosmic abundance at the beginning of the chemical simulations.

\begin{figure}[h] 
    \centering
    \includegraphics[width=1\linewidth]{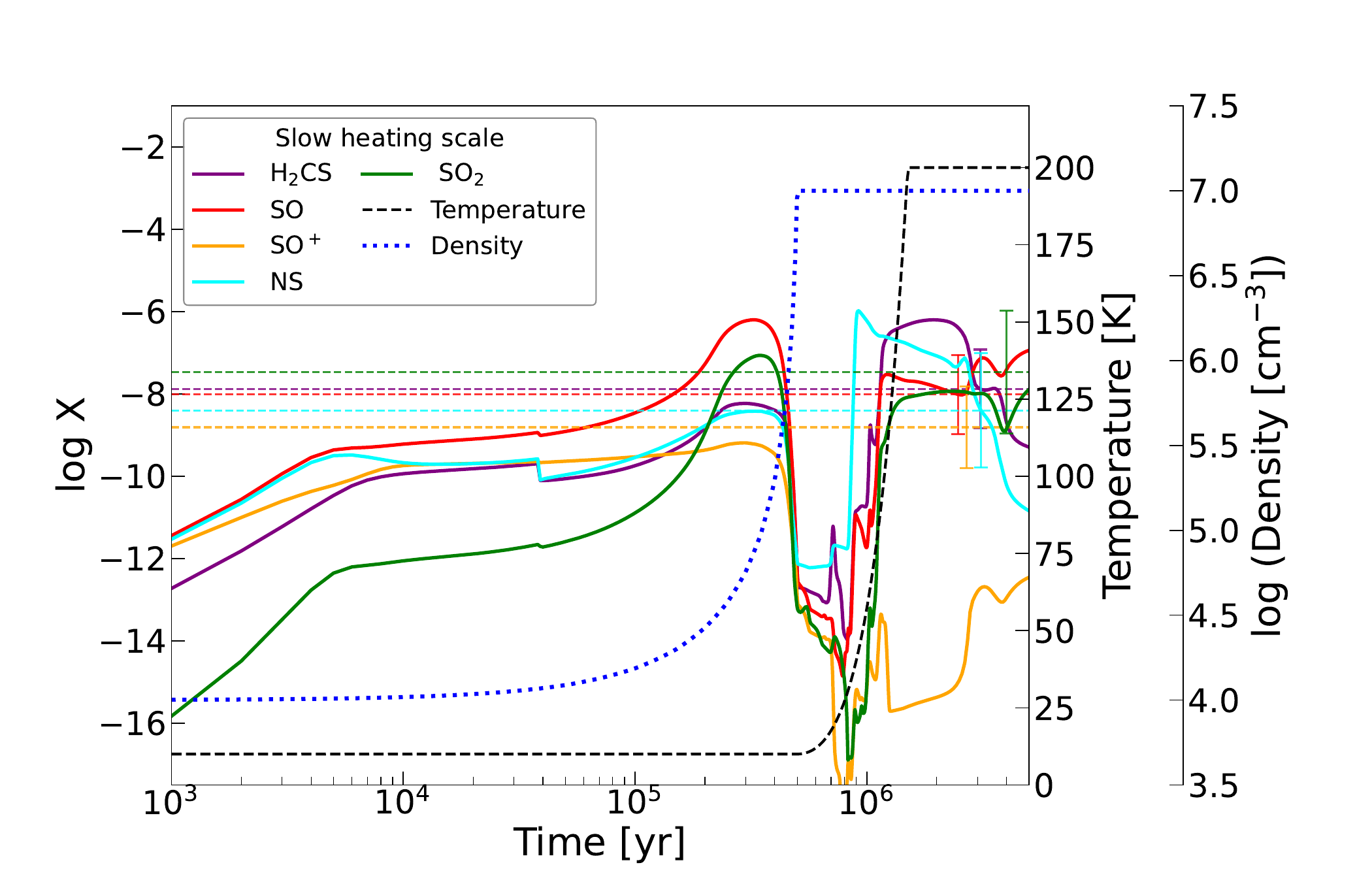}\hfill
    \includegraphics[width=1\linewidth]{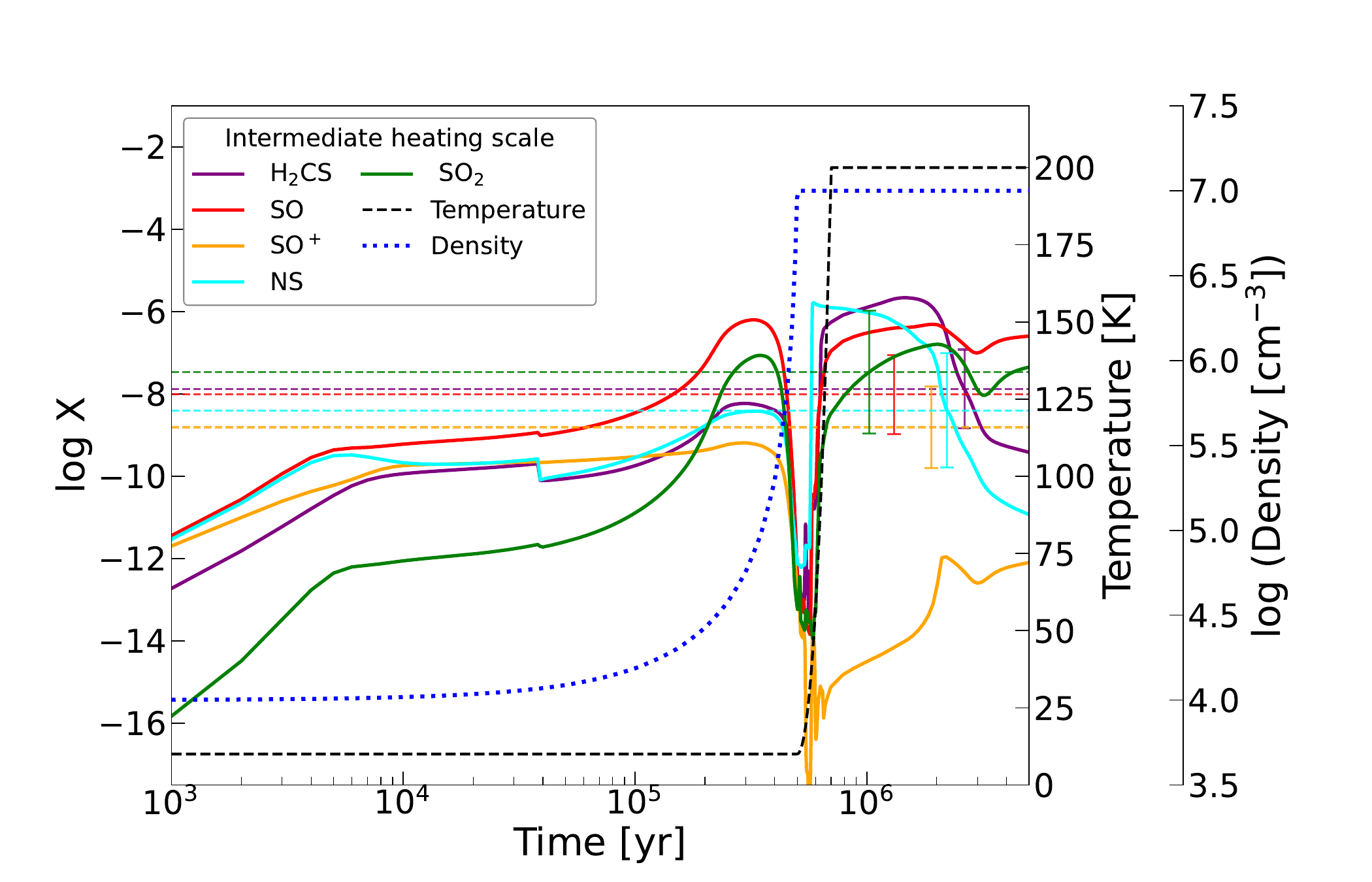}\hfill\\
    \includegraphics[width=1\linewidth]{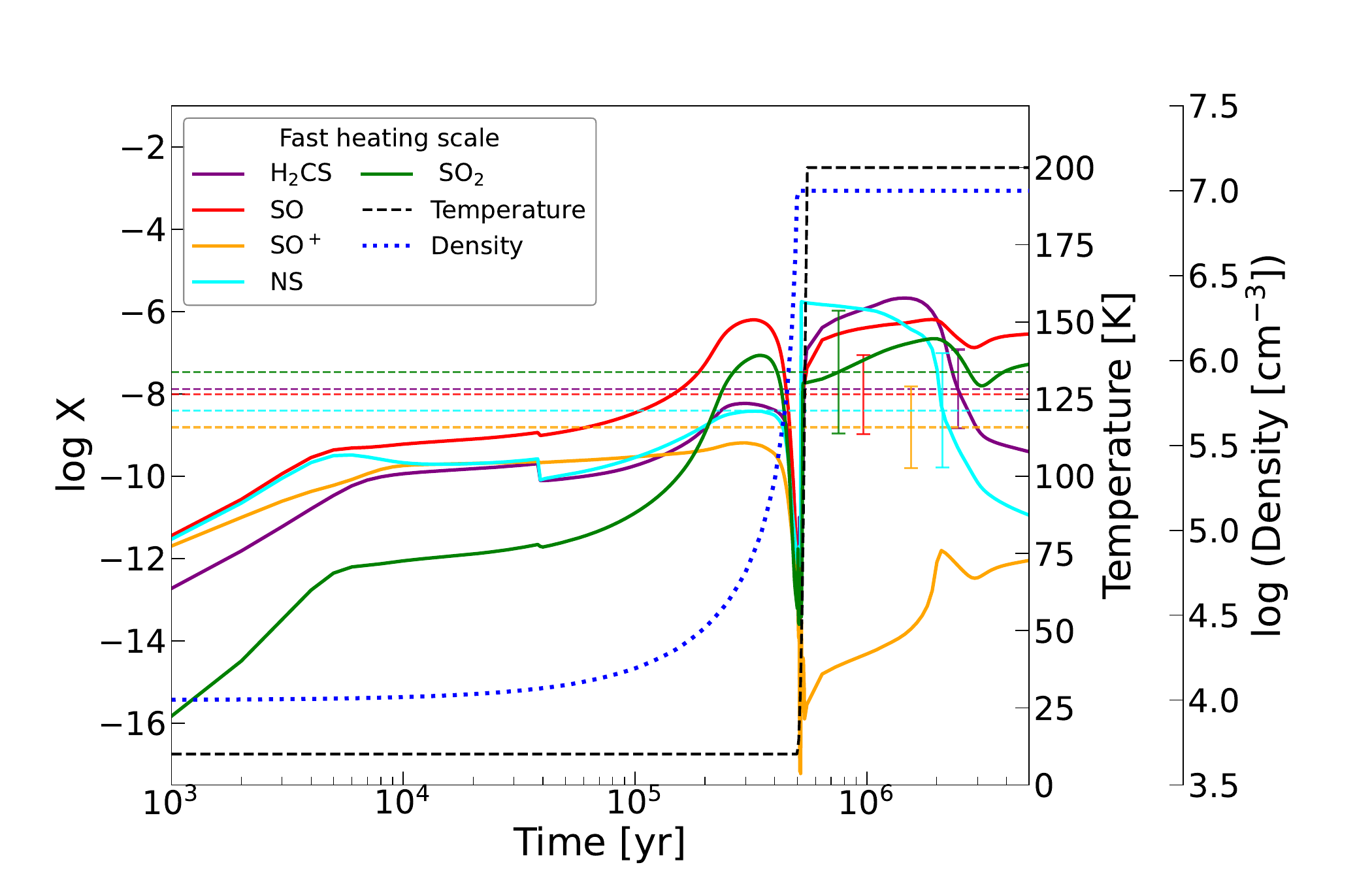}
    \caption{Simulated abundances of SO, SO$_2$, NS, SO$^+$, and H$_2$CS as a function of time. The upper, middle, and bottom panels illustrate the chemical evolution under the slow ($1 \times 10^6$ yr), intermediate ($2 \times 10^5$ yr), and fast ($5 \times 10^4$ yr) warm-up timescales, respectively. In all panels, the dashed black line indicates temperature behaviour, and the dotted blue line illustrates the density increase. The colored horizontal dashed lines represent the mean observational abundances derived for each species across the sample, with the corresponding vertical error bars indicating the 2$\sigma$ uncertainty.}
    \label{nautilus}
\end{figure}

Figure\,\ref{nautilus} presents the modelled fractional abundances of the sulfur-bearing species (except $\rm ^{34}SO$) as a function of time for the slow, intermediate, and fast warm-up scales (upper, middle, and bottom panels, respectively). As depicted in the figure, the cold collapse phase ($\rm t \leq 5 \times 10^5$~yr) is identical across all three scenarios. During the initial stages of this isothermal period, gas-phase chemistry actively synthesises sulfur-bearing molecules from the primordial atomic pool. Species accumulate simultaneously in both the gas and solid phases, albeit at different evolutionary rates. 

As done in \citet{vidal17}, to quantitatively compare our observational results with the outputs of the three above-mentioned model trajectories, the distance of disagreement ($\rm D(t)$) metric was implemented \citep{wakelam06}. This statistical parameter calculates the average logarithmic difference between the observed and modelled abundances for all five analysed sulfur-bearing species simultaneously as:

\begin{equation}
 D(t)=\frac{1}{N_{\rm obs}} \sum_{i}|~log[X_{\rm obs,i}]-log[X_{\rm mod, i}(t)]~|
\end{equation}

\noindent where $X_{\rm obs,i}$ and $X_{\rm mod,i}(t)$ are the observed abundances of the i-species and the modelled ones at time t, respectively. $N_{\rm obs}$ is the total number of observed species considered. By identifying the time at which $D(t)$ reaches its minimum during the warm-up phase, the post-collapse chemical age of each core can be estimated, allowing for an assessment of how well a purely thermal scenario reproduces the observed state of the gas. Crucially, $D(t)$ was calculated independently for each of the 16 cores rather than relying on a sample-wide average. This source-by-source approach is needed because each core may possess unique local conditions and distinct levels of protostellar feedback. Evaluating them individually allows for a systematic distinction between sources governed mainly by thermal heating and those influenced by non-thermal mechanisms—such as outflow-driven turbulence—where passive heating models fail to capture the emission.  


Applying this source-by-source minimisation routine to the sample yields the results summarised in Table\,\ref{dod}. The table reports the absolute minimum distance of disagreement obtained for each of the 16 cores, along with their respective optimal heating model and the corresponding estimated post-collapse chemical age. We consider sources with $D_{\rm min} < 1$ to show a good correlation between observations and models; hence, the resulting times can be considered reliable chemical ages. On the other hand, sources with $D_{\rm min} > 1$ indicate that in these cases the chemistry based on a simple heating model is not well reproduced. This will be further discussed in Sect.\,\ref{modeldiscuss}.

\begin{table}[h!]
\centering
\caption{Results of the $D(t)$ minimization routine. For each source, the preferred heating model, the absolute minimum distance of disagreement ($D_{\rm min}$), and the derived post-collapse chemical age are included.}
\label{dod}
\begin{tabular}{c c c c}
\hline\hline
\# & Preferred & $D_{\rm min}$ & Chemical Age \\ 
   & Model     &           & (yr)          \\ 
\hline
1  & FAST  & $0.14$     & $5.3 \times 10^5$ \\ 
2  & SLOW  & $0.97$     & $4.0 \times 10^6$ \\ 
3  & FAST  & $1.22$   & $2.6 \times 10^6$ \\ 
4  & SLOW  & $0.48$     & $3.9 \times 10^6$ \\ 
5  & FAST  & $1.22$   & $2.7 \times 10^6$ \\ 
6  & FAST  & $0.39$     & $5.3 \times 10^5$ \\ 
7  & FAST  & $1.24$   & $2.5 \times 10^6$ \\ 
8  & SLOW & $0.38$     & $2.9 \times 10^6$ \\ 
9  & FAST  & $1.13$   & $2.3 \times 10^6$ \\ 
10 & INTER.  & $1.07$   & $2.5 \times 10^6$ \\ 
11 & INTER.  & $1.22$   & $2.4 \times 10^6$ \\ 
12 & INTER. & $1.12$   & $2.4 \times 10^6$ \\ 
13 & SLOW  & $1.23$   & $2.9 \times 10^6$ \\ 
14 & SLOW & $1.08$   & $3.1 \times 10^6$ \\ 
15 & SLOW  & $0.39$     & $3.1 \times 10^6$ \\ 
16 & SLOW  & $0.39$     & $3.1 \times 10^6$ \\ 
\hline
\end{tabular}
\end{table}

\section{Discussion}
\label{sectdiscuss}

This section discusses our findings, first focusing on the observational results and then on those from the chemical modelling. Additionally, we explore the implications arising from the comparison between the modelling outcomes and the observations.

\subsection{Interpretation of the observational results}


From the analysis of the abundances ($X$) of the sulfur-bearing molecules in the sample of evolved cores within a temperature range exceeding 100 K, and reaching up to about 220 K in some cases, we first observe relatively low correlations between $X$ and $T$ (see Figs.\,\ref{spearman} and \ref{scatterSulfurT}). As shown in the second figure, the correlations, measured using the Spearman rank correlation coefficient, yield values below or close to 0.5 in most cases, except for H$_2$CS, which shows a stronger correlation with a coefficient of about 0.8. Considering the results reported by \citet{martinez24} for early-stage cores, where molecular abundances were found to be strongly and positively correlated with temperature, our results suggest that this correlation weakens above 100 K for most sulfur-bearing species, except for H$_2$CS. This can be observed in Fig.\,\ref{fig:diff_spearman_abundances} (upper panel), which compares the correlations between both types of samples. This result is consistent with the fact that, once gas-phase chemistry becomes dominant, it can be destructive for certain species or lead to a plateau in their abundance profiles \citep{charnley97,viti01,esplugues14}.

For instance, as pointed out by \citet{charnley97}, sulfur monoxide is rapidly formed from H$_2$S and converted into SO$_2$ within $10^5$ yr when $T$ < 230 K. Above 230 K, most of the reactive oxygen is channelled into H$_2$O, and consequently SO$_2$ production is inhibited. In our case, although we have only one source with $T$ > 230 K, the observed low correlation of the abundances with temperature could be related to this kind of effect. It is worth mentioning that more recent chemical studies, such as \citet{wakelam11} and \citet{esplugues14}, show that SO forms from O+SH, while SO$_{2}$ is mainly produced through the reaction of SO with OH under hot-cores conditions. Likewise, in agreement with these proposed chemical scenarios, a very high correlation is observed between $X$(SO$_2$) and $X$(SO) (see Fig.\,\ref{spearman}), and the SO$_2$/SO ratio increases notably in evolved cores compared to early ones (Fig.\,\ref{fig:ratios_sulfuradas}). This increase can also be attributed to chemistry triggered by the onset of shocks generated in protostellar outflows \citep{esplugues14,burkhardt19,liu25}. In any case, our results provide observational evidence that the $X$(SO$_2$)/$X$(SO) ratio can be used as a reliable chemical clock in star-forming regions (see \citealt{fontani25}). 

In general, the abundances of all the analysed sulfur-bearing molecules, excepting H$_{2}$CS, are higher in the evolved cores than in the early ones (see Fig.\,\ref{fig:abundancias_sulfuradas} and Table\,\ref{tab:sulphur_abundances_early_evolved}). This suggests a general enrichment of sulfur in the gas phase as the molecular core evolves, in agreement with the fact that sulfur-bearing species are strongly time-dependent and that their evolution depends on the heating rate (\citealt{esplugues14}, and references therein).

The abundance of SO and its isotopologue $^{34}$SO show a coherent increase in the evolved cores, both showing enhancement factors of approximately 10. This agreement confirms that the observed trend represents a true enhancement of sulfur-bearing material, ruling out potential line opacity effects or isolated behaviour associated with a single transition. The ion SO$^+$ also shows a significant increase in the evolved cores, with an enhancement factor of about 12. NS shows a more moderate, but still statistically significant enhancement, with a median abundance increase factor of about 4.5. This suggests that NS also participates in the general sulfur enrichment observed in the evolved sample, although its increase is less pronounced than that of the oxygen-containing sulfur species. This may be because abundances of oxygen-bearing species increase more than others, such as carbon- and nitrogen-bearing species, likely due to the greater availability of atomic oxygen during evolution generated by the photodissociation of water \citep{fontani23}.

On the other hand, the abundance of H$_2$CS shows the smallest difference between both evolutionary groups. Although its median abundance is higher in the evolved cores, the enhancement factor is only about 2, and the Mann--Whitney (p)-value, used to evaluate whether the abundance
distributions of two samples are statistically different, is not significant. 
This behaviour is consistent with models presented by \citet{hatch98}. While volatile oxides like SO and SO$_2$ may increase by orders of magnitude in warm environments, the abundance of H$_2$CS remains quite uniform. According to \citet{charnley97}, H$_2$CS can become
very abundant at high temperatures but only after substantial
destruction of H$_2$S. More recent, \citet{esplugues14,esplugues22} proposed that  H$_{2}$CS is mainly formed through the reactant CH$_{3}$ in hot cores as well as in cold cores. For both types of cores, H$_{2}$CS has also been found to be, together with SO$_{2}$, one of the most abundant sulfur-bearing molecules from theoretical and observational results (e.g., \citealt{esplugues22,fernandez26}), which is in close agreement with our findings.

In general, the abundance correlation differences shown in Fig.\,\ref{fig:diff_spearman_abundances} indicate that the transition from early to evolved cores is not simply associated with a uniform strengthening of all correlations. Instead, some molecular relationships become weaker in the evolved sample. In particular, the correlations involving H$_2$CS tend to decrease with respect to several other sulfur-bearing species, such as SO, SO$_2$, SO$^+$, and $^{34}$SO. This suggests that H$_2$CS becomes partially decoupled from the sulfur chemistry involving the other species in the evolved cores. This is in agreement with what was discussed above regarding the striking uniformity of the H$_2$CS abundance, which stands in sharp contrast to the variations in the abundance of the other molecular species between the early and evolved core samples.

\subsubsection{Abundance ratios}

Regarding the presented comparisons between the abundance ratios of the early and evolved core samples (see Fig.\,\ref{fig:ratios_sulfuradas}), the most remarkable differences involve ratios containing SO$_2$. The SO$_{2}$/$X$ ratios are systematically higher in the evolved cores. This indicates that SO$_2$ does not simply increase together with the other sulfur-bearing species, but becomes relatively more dominant in the evolved sample. Overall, oxygen-bearing molecules show an increased abundance relative to oxygen-free species. These results agree with findings in low-mass star-forming regions \citep{esplugues23,fernandez26}. The particularly pronounced increase in the $X$(SO$_{2}$) is consistent with oxygen-bearing chemistry becoming dominant at warm/hot temperatures, favouring SO$_2$ production likely driven not only by its own formation pathways but also by the efficient conversion of SO into SO$_2$ \citep{vidal18,fontani23}.

The abundance ratios comparison reinforces the idea that SO$_2$ plays a central role in the chemical differentiation between early and evolved cores. Its strong absolute enhancement, together with its relative increase with respect to other sulfur-bearing molecules, particularly with SO, supports the interpretation that SO$_2$, and the SO$_{2}$/SO ratio, are particularly sensitive to the physical and chemical changes associated with the evolution of massive star-forming cores \citep{fontani23}. The chemical modelling presented here, which distinguishes between well- and poorly-modelled sources, supports this (see Sect.\,\ref{modeldiscuss}).

\subsubsection{Line-width behaviour and kinematic correlations}

The analysis of the line widths, $\Delta$v, provides  information on the kinematic behaviour of the gas and can be used to
analyse the location or distribution of the molecular species in
different gaseous layers with different dynamics within the cores \citep{sanhueza12,yu15,fontani23}. The comparison between early and evolved cores (see Fig.\,\ref{deltaV_medians}) suggests that the molecular line widths change only slightly between the two types of samples. In particular, the oxygen-bearing species show either small variations or a decrease in their median $\Delta$v. In contrast, the non-oxygen-bearing species, NS and H$_2$CS, which may trace more external layers of the cores \citep{martinez24}, show a slight increase in their median line widths. 

We find that the line-width correlations among the studied sulfur-bearing tracers tighten systematically as cores evolve from the younger, colder phase toward the warmer, protostellar stage. The Spearman correlation coefficients of the $\Delta$v line widths increase from the early to the evolved samples for the majority of molecular pairs, yielding
$\Delta \rho_{\rm \Delta v} > 0$.
In particular, pairs involving key tracers such as SO$_2$ and $^{34}$SO exhibit the most pronounced strengthening of their $\Delta$v correlations, underscoring a progressive coupling of the gas components as star formation advances (see Fig.\,\ref{diffdeltav}).

Physically, this trend can be understood in terms of two complementary processes driving chemical and dynamical homogenization. First, in the early and cold cores, differential depletion and chemical stratification cause molecules to trace distinct gas parcels; as envelopes warm and ice mantles sublimate, sulfur-bearing species are released and mixed within the same gas volume, promoting co-spatiality \citep{esplugues14, vidal17}. Second, the transition to the evolved phase is accompanied by a shift from fragmentation-dominated, locally turbulent kinematics to a regime in which the central protostellar potential, envelope rotation, infall, and collimated outflows impose a global dynamical driver. In this context, multiple tracers respond to the same dominant kinetic motions, yielding tighter line-width correlations as the core becomes chemically and dynamically more homogeneous \citep{bachiller1997, codella2021}.

Nevertheless, the coupling is not universal. A subset of tracer pairs, notably involving SO$^+$, becomes negative in $\Delta \rho_{\rm \Delta v}$. This likely reflects that SO$^+$ can remain localised to ion–neutral shock fronts or regions of enhanced irradiation, preserving a degree of decoupling from the bulk gas even in evolved cores. Such exceptions emphasise that despite a global trend toward homogenization, individual chemical pathways and physical environments imprint residual tracer-specific kinematic signatures.

In sum, the observed enhancement of line-width correlations with evolution provides a robust observational proxy for the chemical and dynamical convergence of dense gas during massive-star formation. This interpretation is consistent with the broader picture of sulfur chemistry evolving toward greater complexity and co-spatiality as protostellar activity intensifies, while retaining traceability of localised deviations tied to specific tracers.

\subsection{Chemistry modelling}
\label{modeldiscuss}

One of the most important outcomes of the chemical modelling is identifying regions where the models closely match the observations, as well as others where observational results cannot be adequately modelled (sources with $D_{\rm min}$<1, and $D_{\rm min}$>1, respectively, presented in Table\,\ref{dod}). For instance, the case of SO$_2$/SO ratio is particularly striking. While the other abundance ratios are well reproduced by the simulations, the modelling fails to reproduce SO$_2$/SO ratios.

\begin{figure}[h]
    \centering
    \includegraphics[width=1\linewidth]{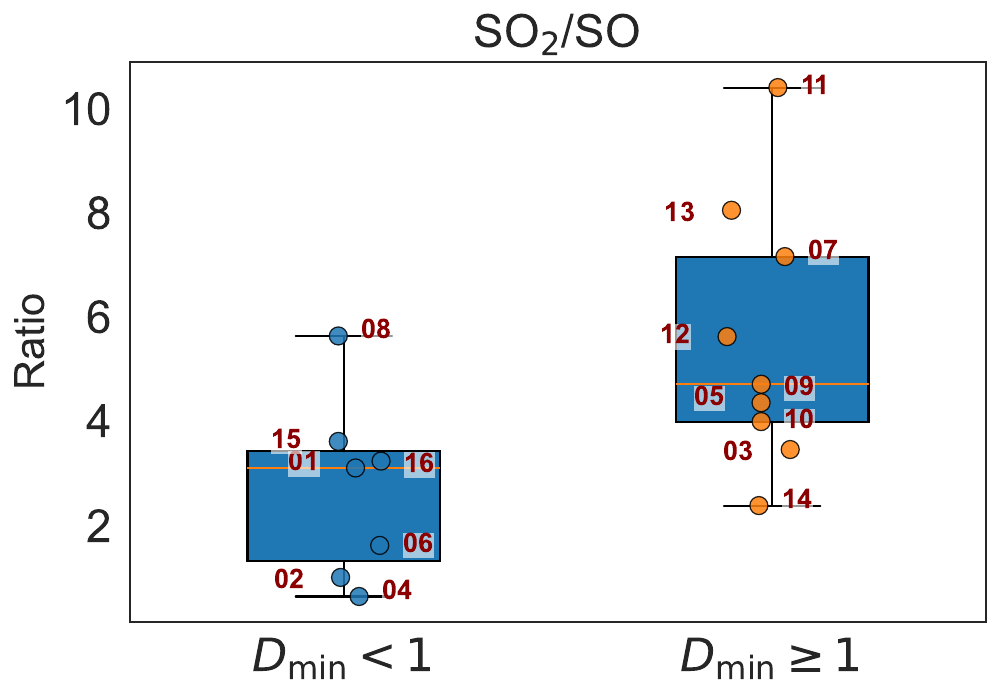}
    \caption{Comparison between the SO$_2$/SO ratios for sources with $D_{\rm min}$ > 1 and < 1 (see Table\,\ref{dod}). The horizontal
line inside the box indicates the median ratio, the box spans
the interquartile range, and the whiskers show the range of non-outlier values. The number of sources is indicated. }
    \label{comparRatio}
\end{figure}

\begin{figure}[h]
    \centering
    \includegraphics[width=1\linewidth]{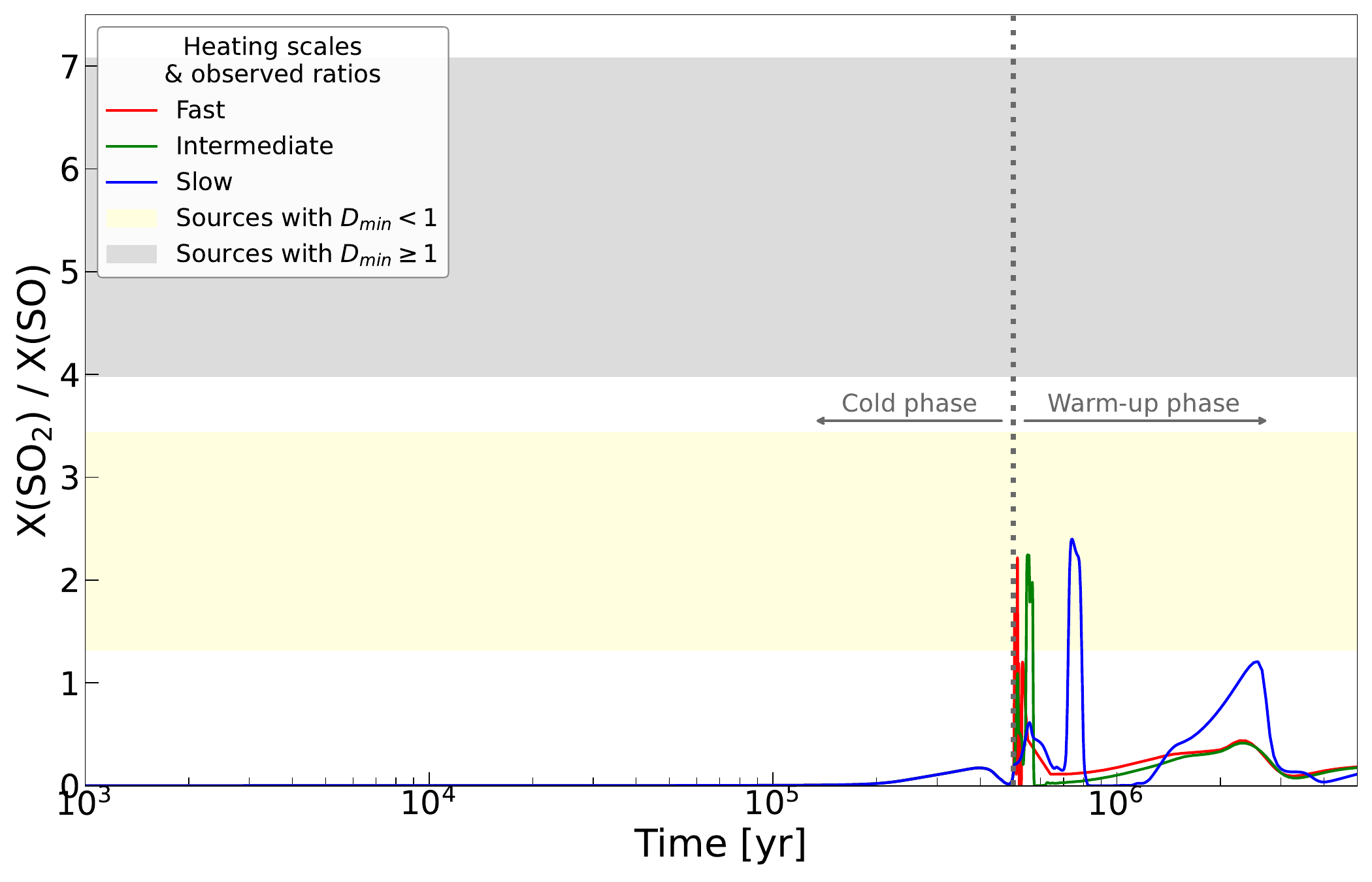}
    \caption{Simulated $X(\text{SO}_2)/X(\text{SO})$ ratio as a function of time. Solid lines illustrate the chemical evolution predicted by the three heating-scale scenarios described in the text. The horizontal coloured-shaded areas indicate the interquartile ranges as presented in Fig.\,\ref{comparRatio}. The vertical dotted line denotes the transition from the initial isothermal cold phase to the warm-up phase.}
    \label{ratio_sim}
\end{figure}

 An interesting difference in the SO$_2$/SO ratio  emerges between sources with  different $D_{\rm min}$. Figure\,\ref{comparRatio} presents the comparison between the SO$_2$/SO ratios of the sources separated by their corresponding
$D_{\rm min}$. We notice that sources with $ D_{\rm min}$ > 1 are those with the highest SO$_2$/SO ratios. Figure\,\ref{ratio_sim}  presents the simulated ratio for SO$_2$/SO, with shaded regions representing the interquartile ranges (as same in Fig.\,\ref{comparRatio}). 
In sources with $ D_{\rm min}$ > 1 (gray-shaded area), models significantly underpredict the observed SO$_2$/SO ratio.  

Considering that our simulations do not account for the presence of shocks induced by molecular outflows, and that the increase in the SO$_2$/SO ratio can be attributed to shock-driven chemistry \citep{esplugues14,burkhardt19,fontani23}, we conclude that this separation among sources arises from their inherent kinematic processes.  In such evolved sources, shock-induced chemistry may be dominant, due to the release of S-bearing molecules from dust grain mantles.
According to \citet{podio15}, between 1\% and 40\% of the elemental S is converted into SO and SO$_{2}$ in the shocks occurring along the jet or at the outflow-cloud interface.


Moreover, the case of SO$^+$ is particularly remarkable. There is only detection of this molecular species in sources that have  $D_{\rm min} > 1$, presumably 
those sources with more pronounced shock-induced kinematic effects. \citet{podio14} point out that this ion, among others, is an effective shock tracer, and they estimated an $X$(SO$^{+}$) of about $ 8 \times 10^{-10}$ in the protostellar shock L1157-B1, which is quite similar to our obtained values.

It is worth mentioning that from an inspection of the $^{12}$CO J=3--2 line contained in the spectral window spw19, we observe that regions with $D_{\rm min} > 1$ generally exhibit more pronounced spectral signatures of shocks and gas kinematics, such as wings and asymmetric self-absorption profiles.
And finally, in this context, it is interesting to add that the same comparison process between observational results and modelling of Sect.\,\ref{models} was carried out for the early cores of \citet{martinez24}; that is, the molecular abundances obtained in that work were compared with the simulations presented here for times prior to the thermal jump. In all cases, values of $D_{\rm min} < 1$ were obtained.

Thus, we conclude that comparing observations and chemical modelling can be useful for distinguishing regions with more pronounced kinematic effects from regions where the gas is still more quiescent. In this regard, chemistry can be used as a tool for probing indirectly the evolution of physical star-formation processes.

\section{Summary and concluding remarks}


Driven by the current astrochemical interest in sulfur and the poorly understood nature of its reservoirs and the formation pathways of sulfur-bearing species, we used interferometric observations to investigate several sulfur-bearing molecules (SO, SO$_2$, SO$^+$, $^{34}$SO, H$_2$CS, and NS)  toward 16 hot molecular cores. This investigation is intended to be an evolutionary continuation of our previous works \citep{martinez24,paron25} in which the same molecules were studied toward a sample of early molecular cores. The main concluding remarks are presented as follows.

We found that the abundances of sulfur-bearing molecules continue to increase with increasing temperature in the analysed range of temperatures, 100 to 220 K. 
The median abundances of all the analysed sulfur-bearing molecules are higher in the evolved cores than in the early ones, suggesting a general
enrichment of sulfur in the gas phase as the molecular core
evolves. This is in agreement with the fact that sulfur-bearing species
are strongly time-dependent and that their evolution depends on
the heating rate. However, we observed that the correlation between abundances and temperature is generally weaker compared to the sample of early sources, suggesting that factors other than temperature, such as gas kinematics, become important in the chemistry of these species.


We found that the abundance ratios, in particular
SO$_2$/$X$ are systematically higher
in the evolved cores than in the earlier ones. This indicates that SO$_{2}$ does not simply increase together with the other sulfur-bearing species, but becomes relatively more dominant in the evolved sample of cores. 

We confirm that the SO$_2$/SO ratio serves as a chemical clock as proposed in the literature. From the comparison between observations and modelling, we found that the evolved sources can be classified into two types: those whose models closely match the observations and those that remain far from them. We propose that the latter have more pronounced kinematic processes, which trigger chemical consequences such as an increase in the SO$_2$/SO ratio.
Thus, we propose that comparing observations with chemical models that omit shock chemistry offers a way to separate more evolved regions—where shocks generated by potential outflows dominate—from less evolved ones. 

From the line-width analysis, we conclude that, in general, due to core evolution and kinematic processes within them, the gas becomes more mixed, losing the notion that different species can trace distinct layers within the cores as found in the early molecular core sample. 

In summary, by tracking the chemical transition from early to evolved molecular cores with high-resolution observations, this study provides key insights into how sulfur-bearing species evolve and behave in the ISM. Our findings demonstrate that the evolution of sulfur-bearing molecules is not merely a linear process of gas-phase enrichment driven by temperature, but rather a complex reorganisation where the analysed species are influenced by gas kinematics. 

\begin{acknowledgements}
We thank the referee for the helpful comments. R.D.T. and N.C.M. are doctoral fellows of CONICET. S.P., A.P., M.E.O., and L.S. are members of the {\it Carrera del Investigador Cient\'\i fico} of CONICET, Argentina. 
This paper makes use of the following ALMA data: ADS/JAO.ALMA\#2022.1.00974.S. ALMA is a partnership of ESO (representing its member states), NSF (USA) and NINS (Japan), together with NRC (Canada), NSTC and ASIAA (Taiwan), and KASI (Republic of Korea), in cooperation with the Republic of Chile. The Joint ALMA Observatory is operated by ESO, AUI/NRAO and NAOJ.

\end{acknowledgements}

%
%

\bibliographystyle{aa}  
\bibliography{ref}
\IfFileExists{\jobname.bbl}{}
{\typeout{}
\typeout{****************************************************}
\typeout{****************************************************}
\typeout{** Please run "bibtex \jobname" to optain}
\typeout{** the bibliography and then re-run LaTeX}
\typeout{** twice to fix the references!}
\typeout{****************************************************}
\typeout{****************************************************}
\typeout{}
}
\label{lastpage}

\begin{appendix}

\section{Example of core emission and spectral line profiles}
\label{example}

\begin{figure}[h]
    \centering
    \includegraphics[width=1\linewidth]{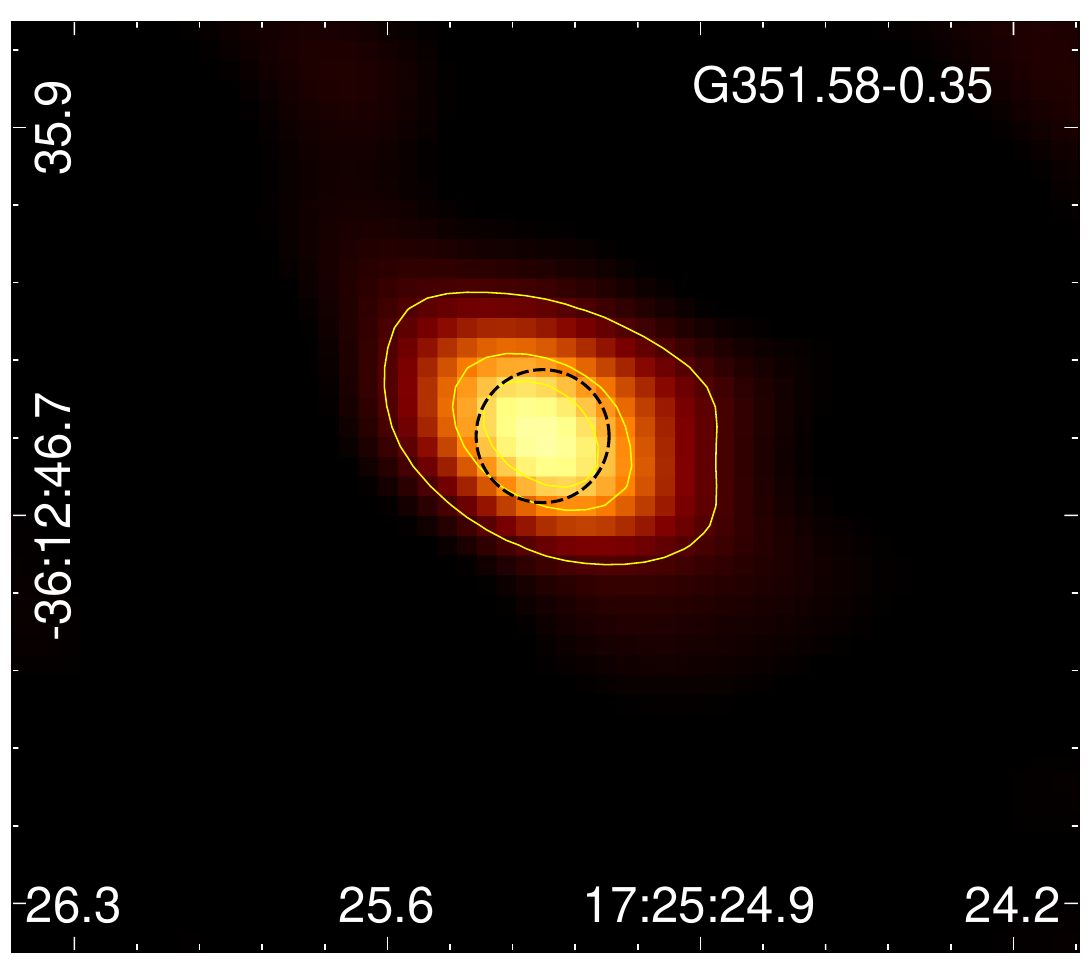}
    \caption{Continuum emission at ALMA Band 7 toward region G351.58-0.35 (core 12). Contour levels are 1, 3, and 4 Jy beam$^{-1}$. The black dashed circle is the beam-size region from which spectra of Fig.\,\ref{spwGauss} were extracted. }
    \label{coreexample}
\end{figure}

\begin{figure}[h!]
    \centering
    \includegraphics[width=0.85\linewidth]{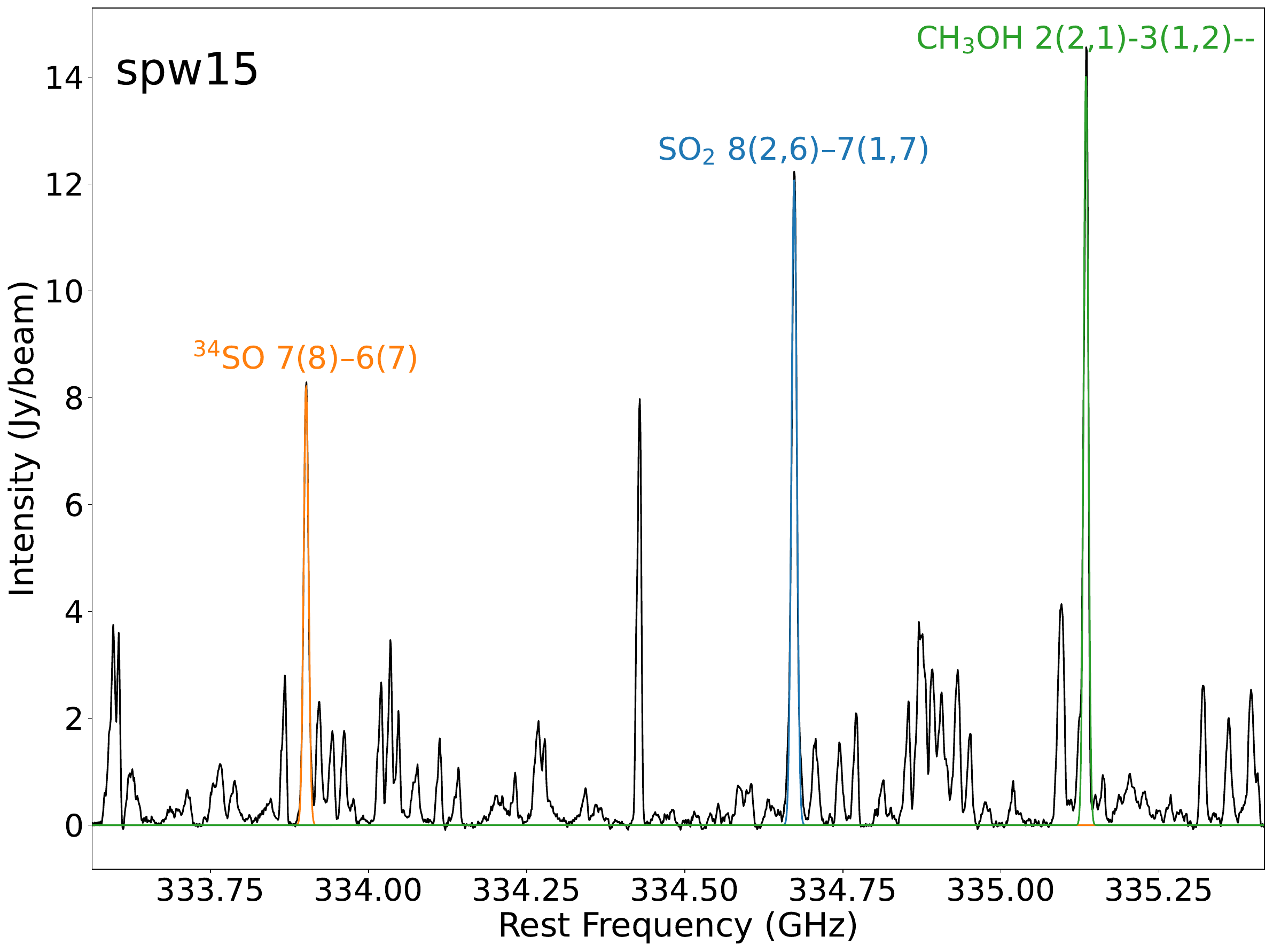}
    \includegraphics[width=0.85\linewidth]{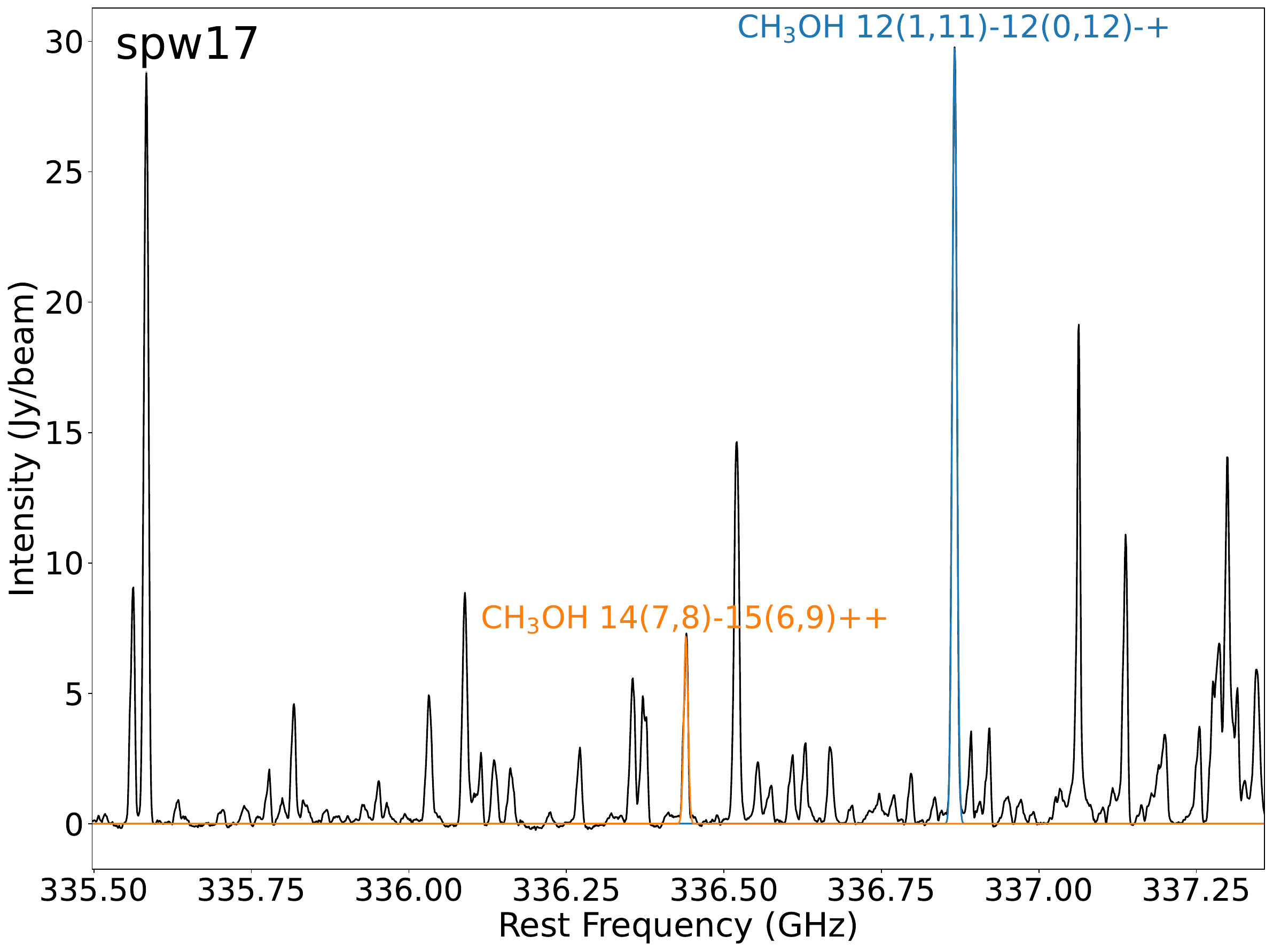}
    \includegraphics[width=0.85\linewidth]{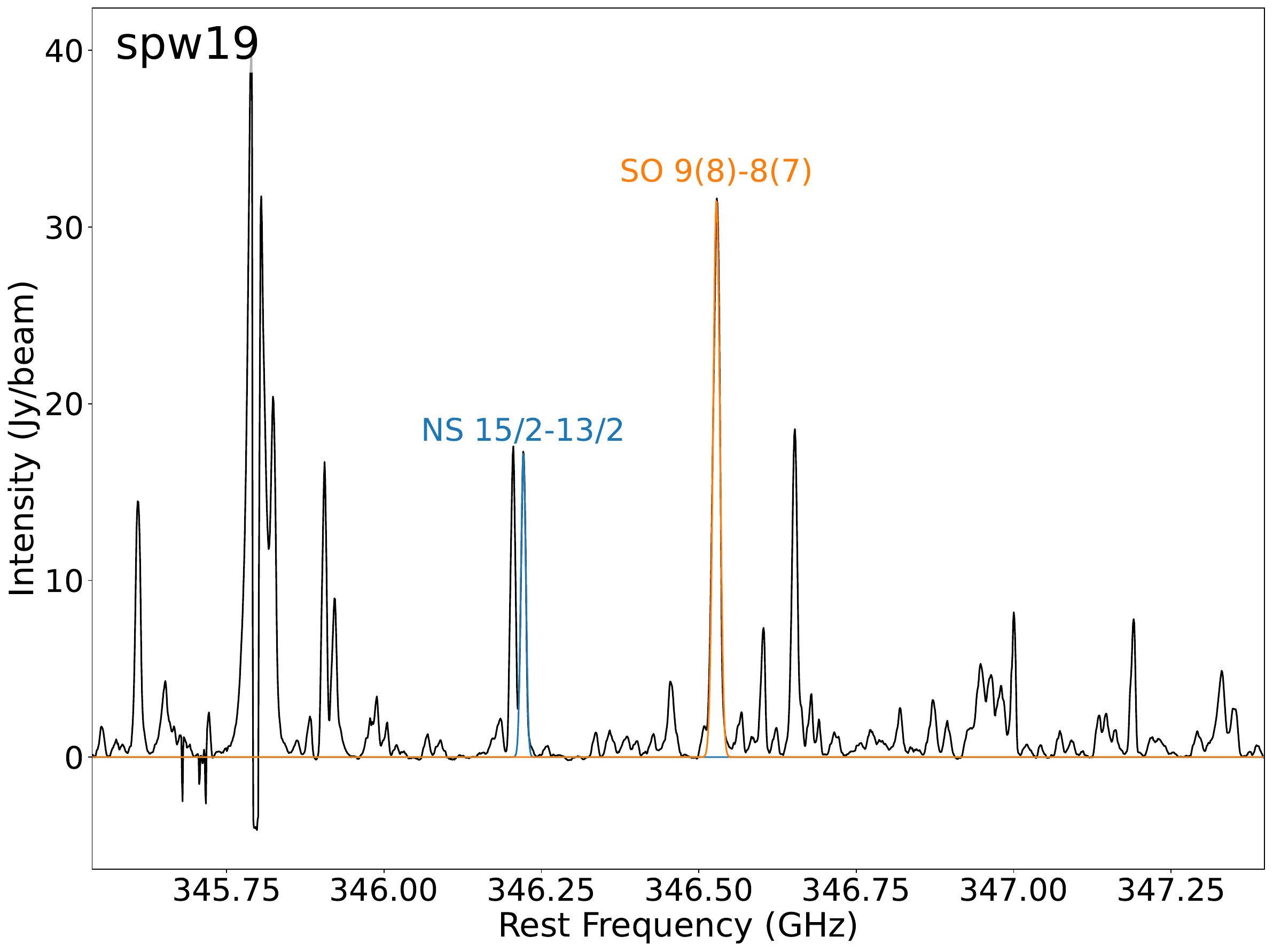}
    \includegraphics[width=0.85\linewidth]{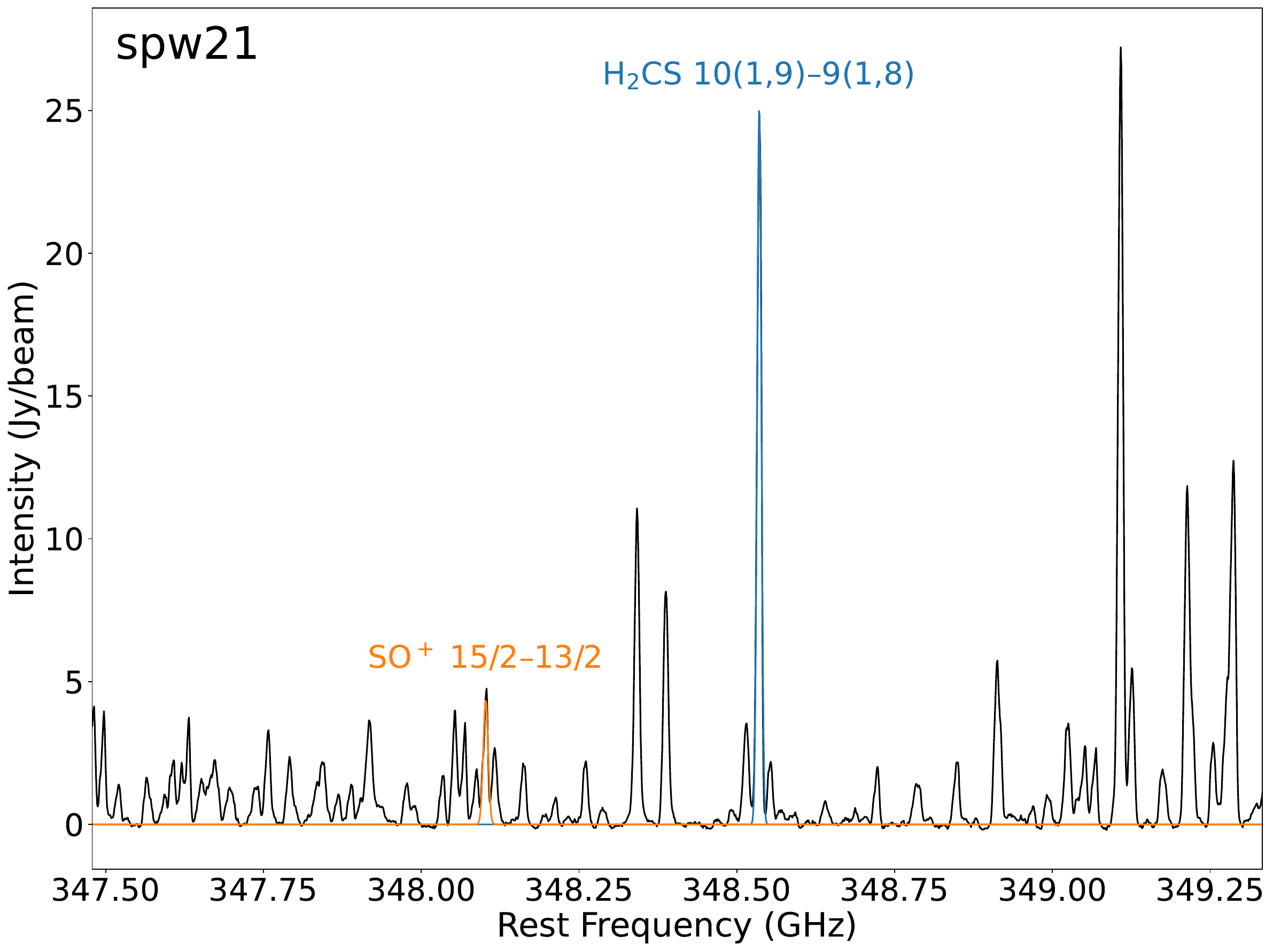}
    \caption{Spectra of the four analyzed spectral windows (spw) obtained from region G351.58-0.35 (core 12). The analyzed lines, sulfur-bearing molecules and methanol transitions, are indicated and gaussian fits are included.}
    \label{spwGauss}
\end{figure}

As an illustrative example of the process carried out for the molecular line analysis, Fig.\,\ref{coreexample} shows the continuum emission at ALMA Band  7 of the G351.58-0.35 region, where core 12 was identified. It is included the beam-size circle at the emission peak from which the spectra presented in Fig.\,\ref{spwGauss} were extracted.

\section{Measured parameters from the molecular lines}
\label{appendGauss}

Tables\,\ref{gauss1}, \ref{gauss2}, and \ref{gauss3} present the line parameters, Peak, $\Delta$v and the integrated emission ($W$), obtained from each analyzed molecular line from Gaussian fittings.

\begin{table*}[h]
\centering
\caption{Parameters obtained from Gaussian fittings for NS and SO.}
\label{gauss1}
\tiny
\begin{tabular}{l|cccccc|cccccc}
\hline
 & \multicolumn{6}{|c|}{NS 15/2--13/2} &   \multicolumn{6}{|c}{SO v=0 $^{3}$$\Sigma$ 9(8)--8(7)} \\
 \hline
\#&Peak&err.&$\Delta$v &err.&$W$&err.&Peak&err.&$\Delta$v&err.&$W$&err. \\
&\multicolumn{2}{c}{\tiny(Jy beam$^{-1}$)}&\multicolumn{2}{c}{\tiny(km s$^{-1}$)} &\multicolumn{2}{c|}{\tiny(Jy beam$^{-1}$ km s$^{-1}$)} &\multicolumn{2}{c}{\tiny(Jy beam$^{-1}$)}&\multicolumn{2}{c}{\tiny(km s$^{-1}$)} &\multicolumn{2}{c}{\tiny(Jy beam$^{-1}$ km s$^{-1}$)} \\
\hline
1 & - & - & - & - & - & - & 1.52 & 0.06 & 5.17 & 0.24 & 8.38 & 0.36 \\
2 & 0.69 & 0.03 & 7.03 & 0.42 & 5.20 & 0.28 & 8.15 & 0.27 & 6.63 & 0.26 & 57.57 & 2.13 \\
3 & - & - & - & - & - & - & 9.08 & 0.28 & 7.23 & 0.26 & 69.94 & 2.39 \\
4 & 0.23 & 0.02 & 3.39 & 0.34 & 0.83 & 0.07 & 2.82 & 0.11 & 3.09 & 0.14 & 9.29 & 0.40 \\
5 & - & - & - & - & - & - & 7.18 & 0.18 & 8.11 & 0.24 & 62.06 & 1.74 \\
6 & - & - & - & - & - & - & 4.09 & 0.14 & 7.17 & 0.29 & 31.31 & 1.19 \\
7 & - & - & - & - & - & - & 6.61 & 0.14 & 7.94 & 0.29 & 39.04 & 1.35 \\
8 & 6.14 & 0.10 & 6.78 & 0.13 & 44.35 & 0.81 & 16.89 & 0.38 & 10.70 & 0.27 & 192.56 & 4.69 \\
9 & 0.78 & 0.02 & 6.08 & 0.23 & 5.10 & 0.18 & 7.24 & 0.24 & 7.41 & 0.28 & 57.22 & 2.08 \\
10 & 1.64 & 0.04 & 4.46 & 0.13 & 7.84 & 0.22 & 5.94 & 0.12 & 6.08 & 0.15 & 38.50 & 0.89 \\
11 & 21.43 & 0.81 & 8.91 & 0.39 & 203.38 & 8.43 & 32.22 & 0.84 & 10.19 & 0.31 & 349.56 & 9.97 \\
12 & 14.22 & 0.17 & 7.71 & 0.11 & 116.72 & 1.53 & 26.61 & 0.52 & 10.92 & 0.24 & 309.42 & 6.58 \\
13 & 32.17 & 1.21 & 13.09 & 0.62 & 448.47 & 19.39 & 41.03 & 1.35 & 16.92 & 0.64 & 739.48 & 26.48 \\
14 & 1.42 & 0.05 & 5.80 & 0.26 & 8.79 & 0.37 & 2.74 & 0.09 & 6.12 & 0.35 & 17.90 & 0.87 \\
15 & 2.13 & 0.05 & 8.49 & 0.24 & 19.33 & 0.50 & 5.78 & 0.13 & 10.00 & 0.26 & 61.63 & 1.54 \\
16 & 2.40 & 0.07 & 5.72 & 0.20 & 14.64 & 0.48 & 8.06 & 0.33 & 7.50 & 0.36 & 64.42 & 2.91 \\
\hline
\multicolumn{9}{l}{A `-' means that it was not observed emission from such molecular line.} \\ 
\end{tabular}
\end{table*}

\begin{table*}[h]
\centering
\caption{Parameters obtained from Gaussian fittings for $^{34}$SO and SO$_{2}$.}
\label{gauss2}
\tiny
\begin{tabular}{l|cccccc|cccccc}
\hline
 & \multicolumn{6}{|c|}{$^{34}$SO 7(8)--6(7)} & \multicolumn{6}{|c}{SO$_{2}$ v=0 8(2,6)--7(1,7)} \\
 \hline
\# &Peak&err.&$\Delta$v &err.&$W$&err.&Peak&err.&$\Delta$v&err.&$W$&err. \\
&\multicolumn{2}{c}{\tiny(Jy beam$^{-1}$)}&\multicolumn{2}{c}{\tiny(km s$^{-1}$)} &\multicolumn{2}{c|}{\tiny(Jy beam$^{-1}$ km s$^{-1}$)} & \multicolumn{2}{c}{\tiny(Jy beam$^{-1}$)}&\multicolumn{2}{c}{\tiny(km s$^{-1}$)} &\multicolumn{2}{c}{\tiny(Jy beam$^{-1}$ km s$^{-1}$)} \\
\hline
1 & 0.14 & 0.03 & 4.37 & 1.21 & 0.66 & 0.16 & 0.27 & 0.04 & 3.81 & 0.62 & 1.08 & 0.15 \\
2 & 0.69 & 0.02 & 7.86 & 0.28 & 5.77 & 0.18 & 0.98 & 0.02 & 7.61 & 0.21 & 7.96 & 0.19 \\
3 & 1.56 & 0.02 & 5.84 & 0.10 & 9.69 & 0.14 & 2.14 & 0.02 & 5.89 & 0.07 & 13.41 & 0.13 \\
4 & 0.17 & 0.03 & 1.85 & 0.40 & 0.34 & 0.06 & 0.32 & 0.03 & 2.39 & 0.38 & 0.81 & 0.14 \\
5 & 1.53 & 0.04 & 7.31 & 0.23 & 11.87 & 0.32 & 1.86 & 3.72 & 7.59 & 1.75 & 15.02 & 0.30 \\
6 & 0.39 & 0.04 & 5.91 & 0.77 & 2.44 & 0.28 & 0.79 & 0.04 & 7.16 & 0.38 & 6.03 & 0.28 \\
7 & 0.99 & 0.03 & 6.50 & 0.22 & 6.82 & 0.20 & 2.17 & 0.03 & 5.93 & 0.08 & 13.70 & 0.16 \\
8 & 2.94 & 0.35 & 7.64 & 1.06 & 23.87 & 2.86 & 5.36 & 0.16 & 9.09 & 0.31 & 51.82 & 1.51 \\
9 & 1.39 & 0.07 & 5.67 & 0.32 & 8.40 & 0.42 & 2.31 & 0.07 & 5.31 & 0.19 & 13.07 & 0.41 \\
10 & 0.69 & 0.03 & 5.66 & 0.33 & 4.14 & 0.21 & 1.35 & 0.05 & 5.24 & 0.23 & 7.54 & 0.29 \\
11 & 11.30 & 2.47 & 9.96 & 2.51 & 119.78 & 26.15 & 12.22 & 1.95 & 10.54 & 2.24 & 137.13 & 23.88 \\
12 & 6.11 & 0.27 & 7.99 & 0.41 & 51.97 & 2.31 & 8.80 & 0.16 & 8.86 & 0.19 & 83.00 & 1.50 \\
13 & 19.57 & 1.56 & 9.08 & 0.84 & 189.20 & 15.08 & 25.28 & 0.71 & 10.94 & 0.36 & 294.40 & 8.29 \\
14 & 0.40 & 0.20 & 4.32 & 3.97 & 1.85 & 1.80 & 0.46 & 0.14 & 3.54 & 1.70 & 1.74 & 0.88 \\
15 & 0.52 & 0.02 & 10.69 & 0.45 & 5.88 & 0.21 & 0.80 & 0.04 & 12.98 & 0.66 & 11.10 & 0.49 \\
16 & 0.81 & 0.06 & 5.73 & 0.49 & 4.92 & 0.37 & 1.49 & 0.03 & 7.12 & 0.18 & 11.30 & 0.24 \\
\hline
\multicolumn{9}{l}{A `-' means that it was not observed emission from such molecular line.} \\ 
\end{tabular}
\end{table*}

\begin{table*}[h]
\centering
\caption{Parameters obtained from Gaussian fittings for SO$^{+}$ and H$_{2}$CS.}
\label{gauss3}
\tiny
\begin{tabular}{l|cccccc|cccccc}
\hline
 & \multicolumn{6}{|c|}{SO$^{+}$ 15/2--13/2 (1/2) l=f } &   \multicolumn{6}{|c}{H$_{2}$CS 10(1,9)--9(1,8) } \\
 \hline
\# &Peak&err.&$\Delta$v &err.&$W$&err.&Peak&Error&$\Delta$v&err.&$W$&err. \\
&\multicolumn{2}{c}{\tiny(Jy beam$^{-1}$)}&\multicolumn{2}{c}{\tiny(km s$^{-1}$)} &\multicolumn{2}{c|}{\tiny(Jy beam$^{-1}$ km s$^{-1}$)} &\multicolumn{2}{c}{\tiny(Jy beam$^{-1}$)}&\multicolumn{2}{c}{\tiny(km s$^{-1}$)} &\multicolumn{2}{c}{\tiny(Jy beam$^{-1}$ km s$^{-1}$)} \\
\hline
1 & - & - & - & - & - & - & 0.53 & 0.01 & 4.32 & 0.14 & 2.44 & 0.07 \\
2 & 0.28 & 0.09 & 6.89 & 0.23 & 2.08 & 0.06 & 2.85 & 0.12 & 5.77 & 0.28 & 17.53 & 0.79 \\
3 & 0.25 & 0.01 & 7.51 & 0.39 & 2.03 & 0.10 & 0.31 & 0.01 & 4.86 & 0.21 & 1.60 & 0.06 \\
4 & - & - & - & - & - & - & 1.30 & 0.02 & 2.73 & 0.07 & 3.78 & 0.09 \\
5 & 0.76 & 0.03 & 7.01 & 0.34 & 5.73 & 0.25 & 1.10 & 0.03 & 5.32 & 0.21 & 6.21 & 0.23 \\
6 & - & - & - & - & - & - & 0.64 & 0.03 & 6.97 & 0.43 & 4.81 & 0.27 \\
7 & 0.22 & 0.01 & 7.94 & 0.50 & 1.90 & 0.11 & 0.43 & 0.01 & 5.77 & 0.19 & 2.67 & 0.08 \\
8 & - & - & - & - & - & - & 11.32 & 0.19 & 6.21 & 0.12 & 74.88 & 1.37 \\
9 & 0.38 & 0.02 & 5.54 & 0.34 & 2.25 & 0.13 & 2.44 & 0.05 & 3.78 & 0.09 & 9.87 & 0.22 \\
10 & 0.19 & 0.01 & 7.08 & 0.52 & 1.48 & 0.09 & 5.60 & 0.13 & 4.26 & 0.11 & 25.42 & 0.65 \\
11 & 4.56 & 0.25 & 13.50 & 1.33 & 65.66 & 5.32 & 34.97 & 0.64 & 7.10 & 0.15 & 264.54 & 5.29 \\
12 & 2.09 & 0.02 & 8.35 & 0.14 & 18.62 & 0.28 & 20.65 & 0.34 & 6.80 & 0.13 & 149.72 & 2.67 \\
13 & 5.67 & 0.10 & 8.76 & 0.20 & 52.93 & 1.10 & 22.39 & 0.48 & 10.49 & 0.27 & 250.04 & 5.98 \\
14 & 0.24 & 0.01 & 12.72 & 1.46 & 3.16 & 0.30 & 3.70 & 0.16 & 4.64 & 0.23 & 18.32 & 0.85 \\
15 & - & - & - & - & - & - & 5.26 & 0.14 & 6.83 & 0.21 & 38.27 & 1.14 \\
16 & - & - & - & - & - & - & 6.22 & 0.15 & 4.76 & 0.13 & 31.58 & 0.83 \\
\hline
\multicolumn{9}{l}{A `-' means that it was not observed emission from such molecular line.} \\ 
\end{tabular}
\end{table*}

\clearpage

\section{Calculation of $T_{\rm rot}$ from the CH$_{3}$OH}
\label{appmetanol}

\begin{figure*}[h!]
    \centering
    \includegraphics[width=0.24\linewidth]{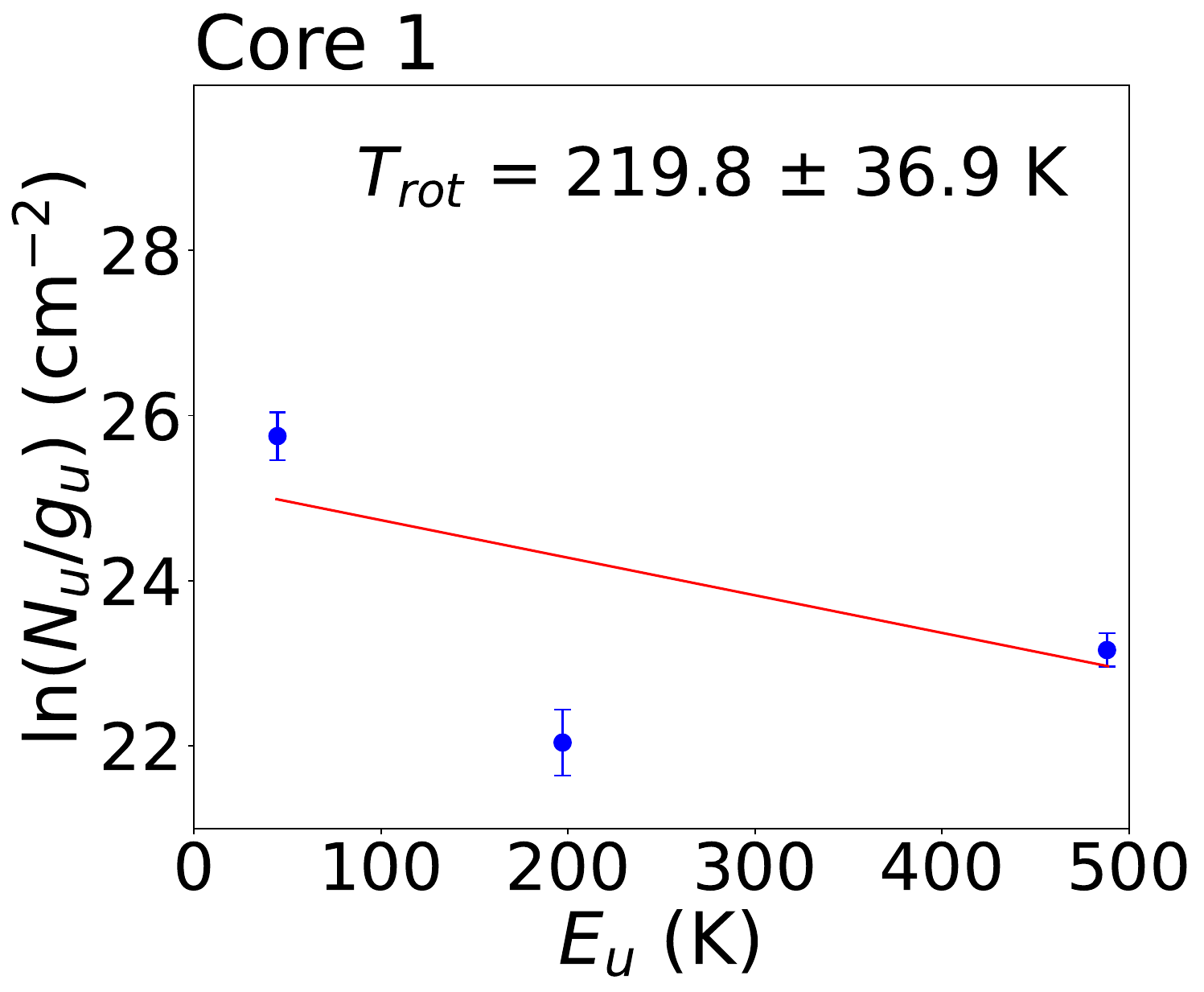}
    \includegraphics[width=0.24\linewidth]{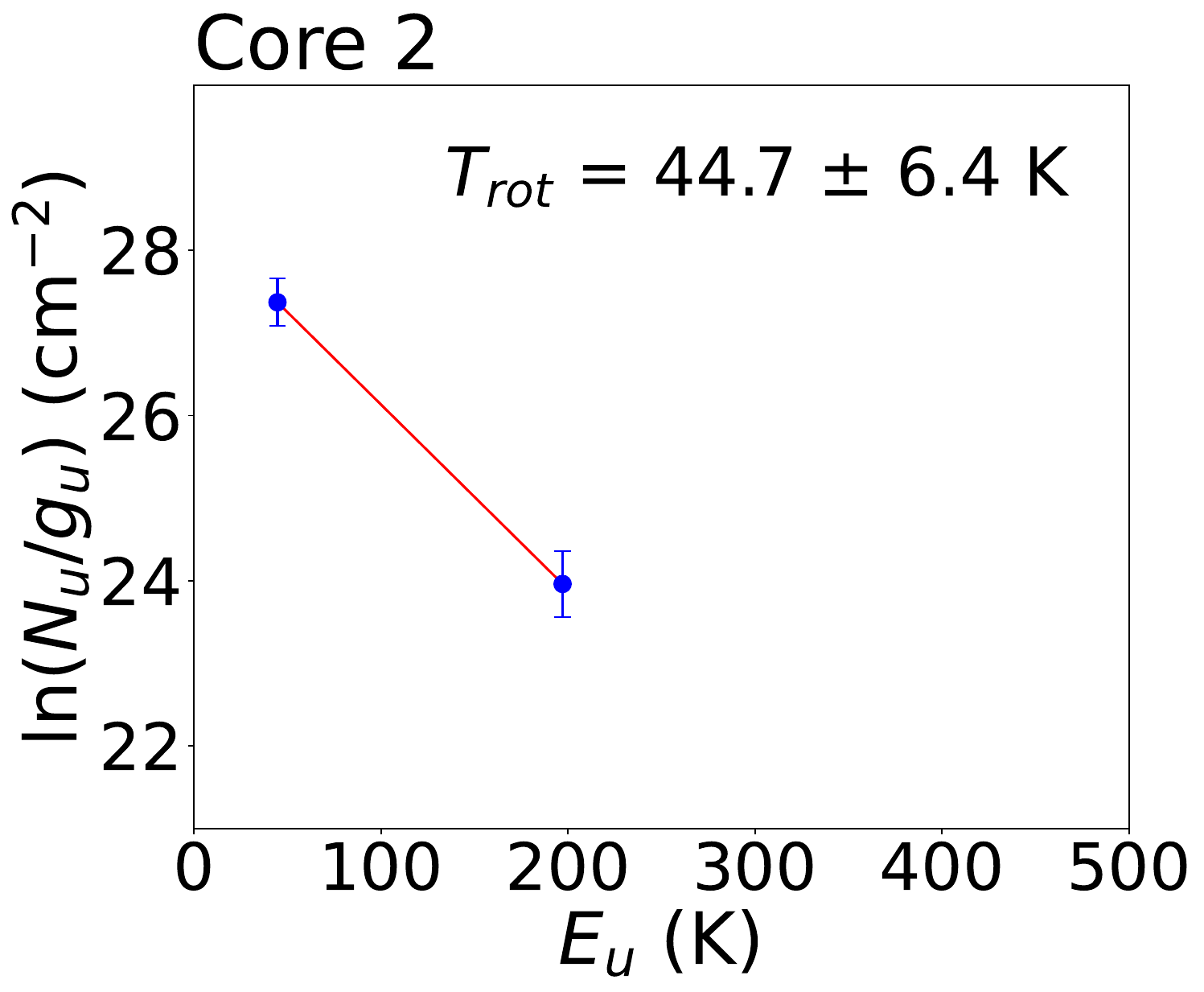}
    \includegraphics[width=0.24\linewidth]{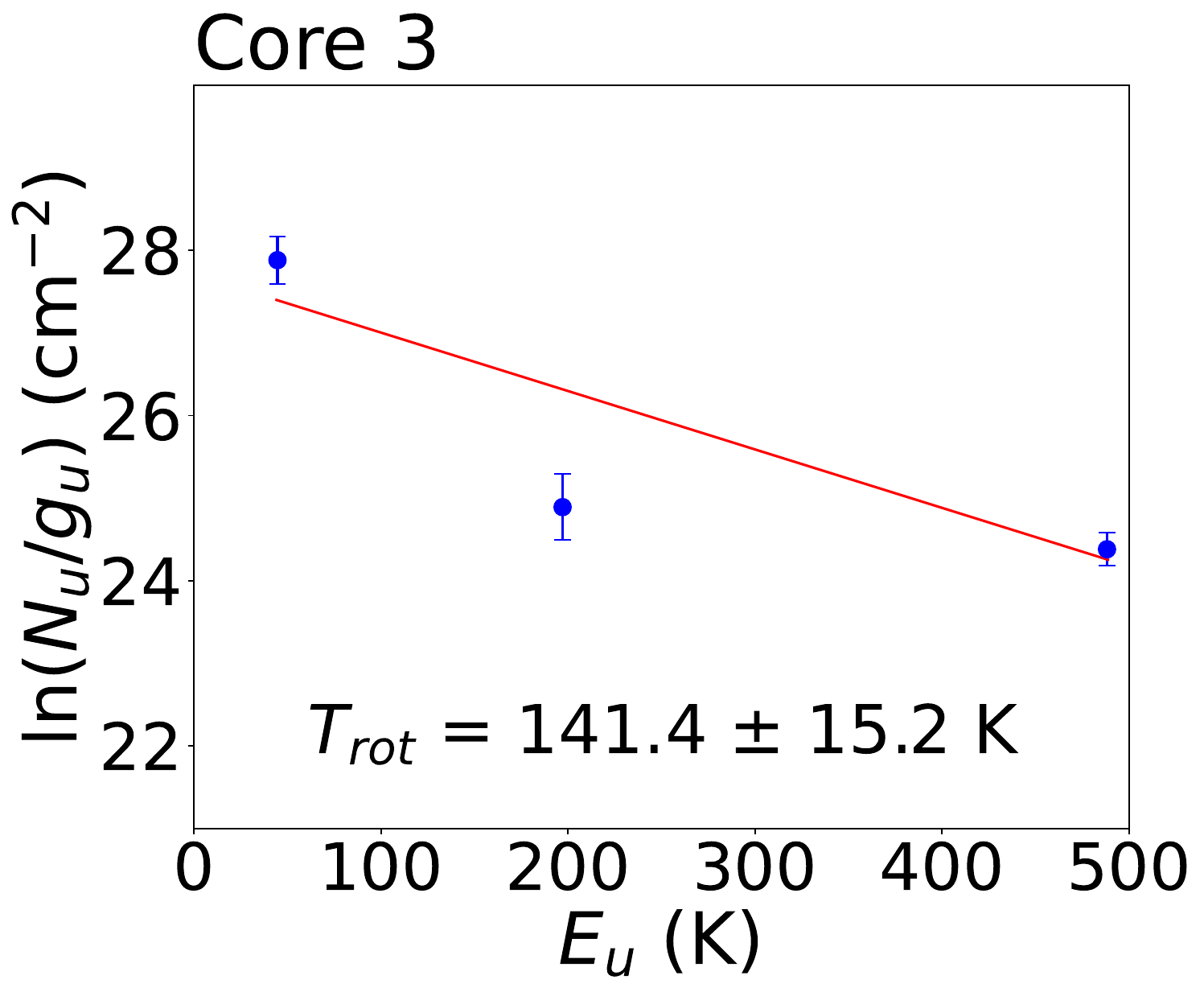}
   \includegraphics[width=0.24\linewidth]{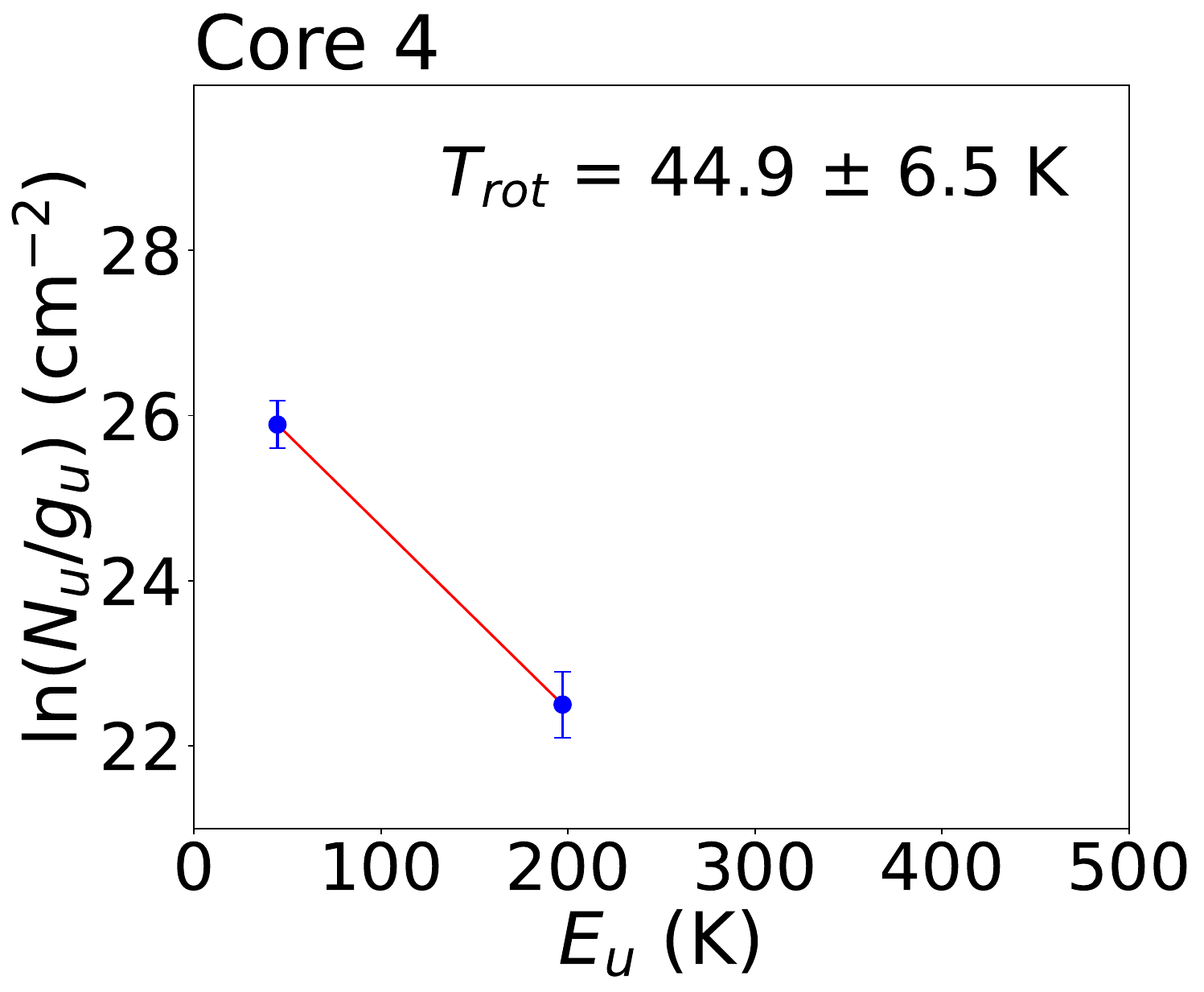}
   \includegraphics[width=0.24\linewidth]{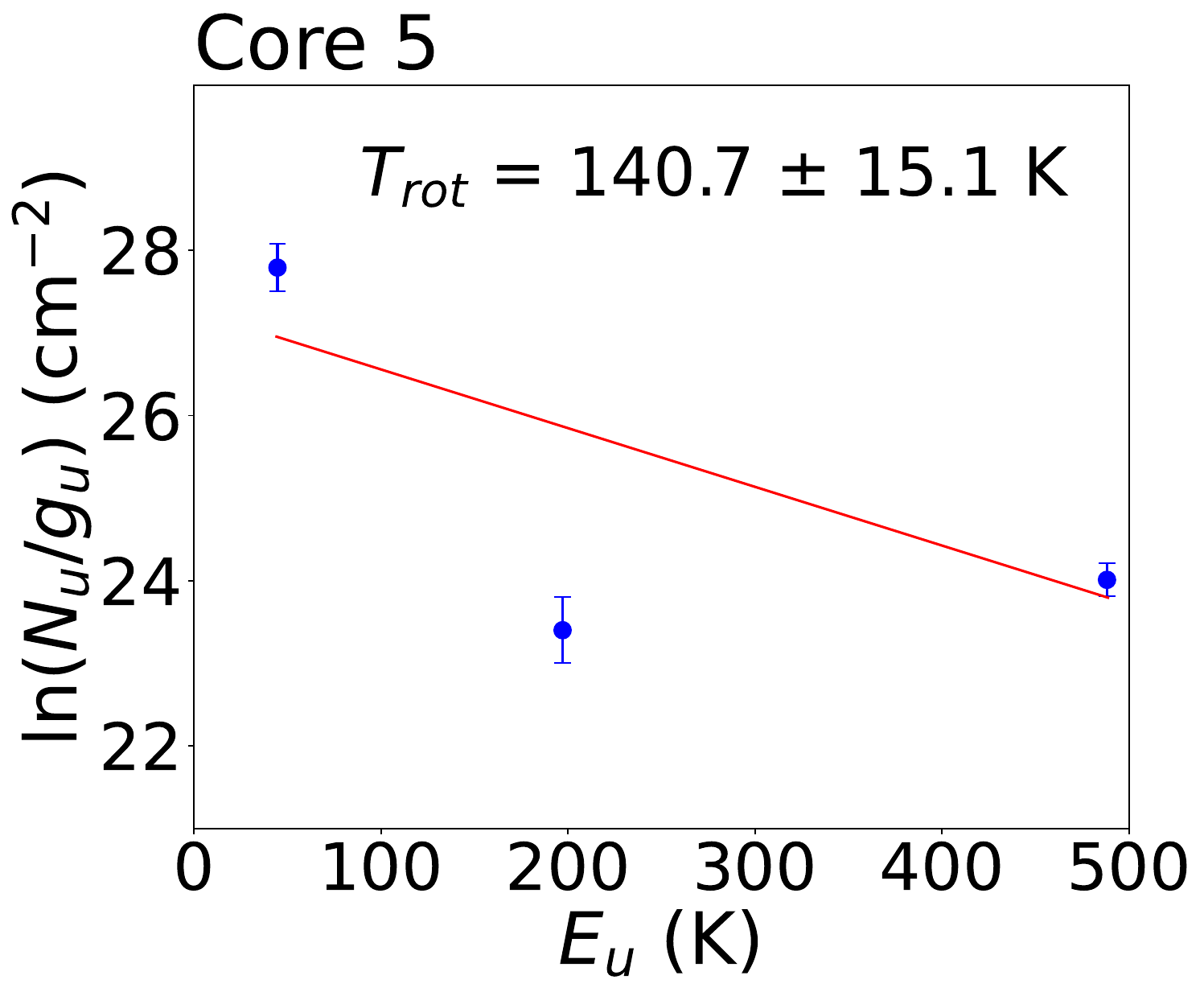}
   \includegraphics[width=0.24\linewidth]{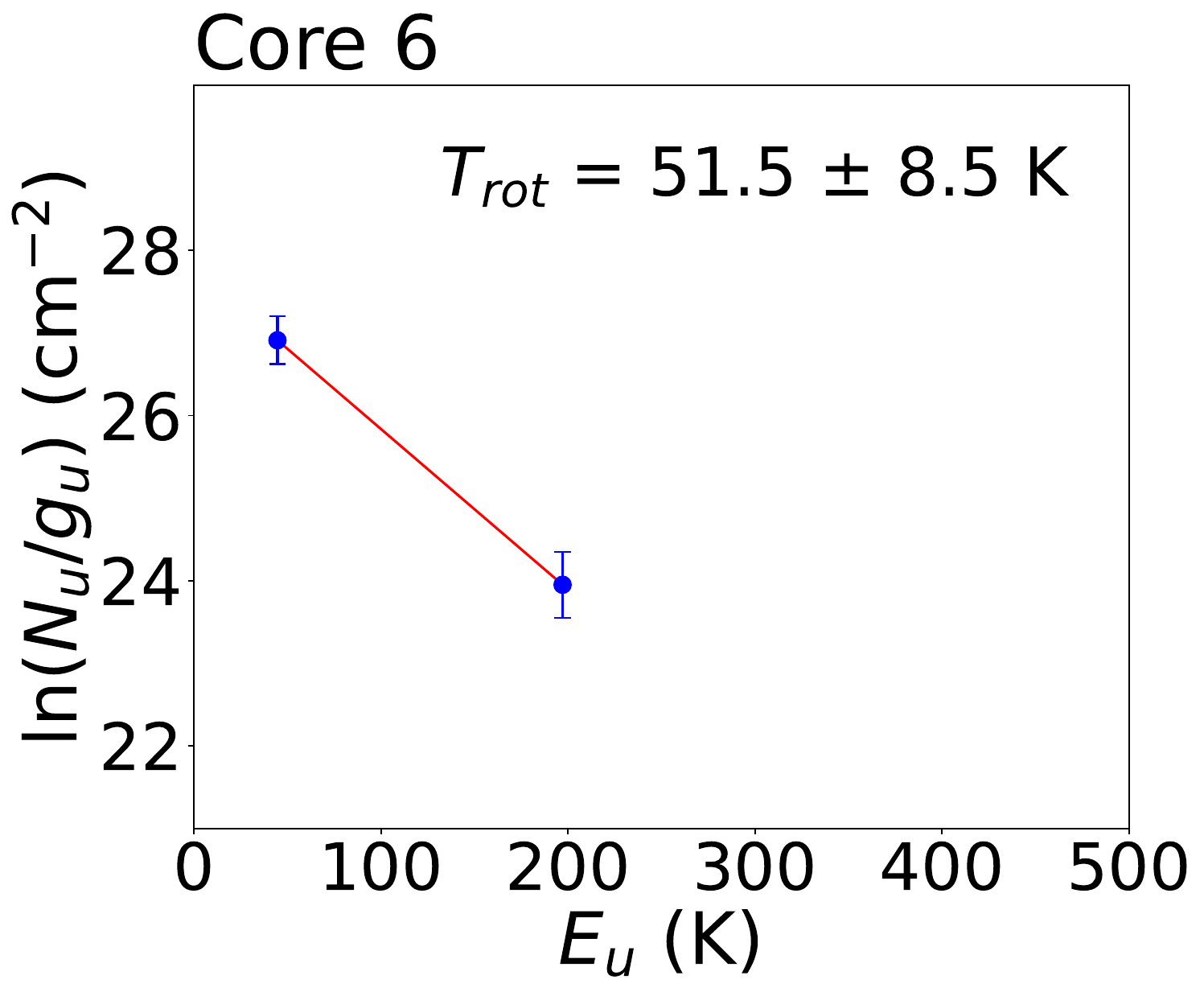}
   \includegraphics[width=0.24\linewidth]{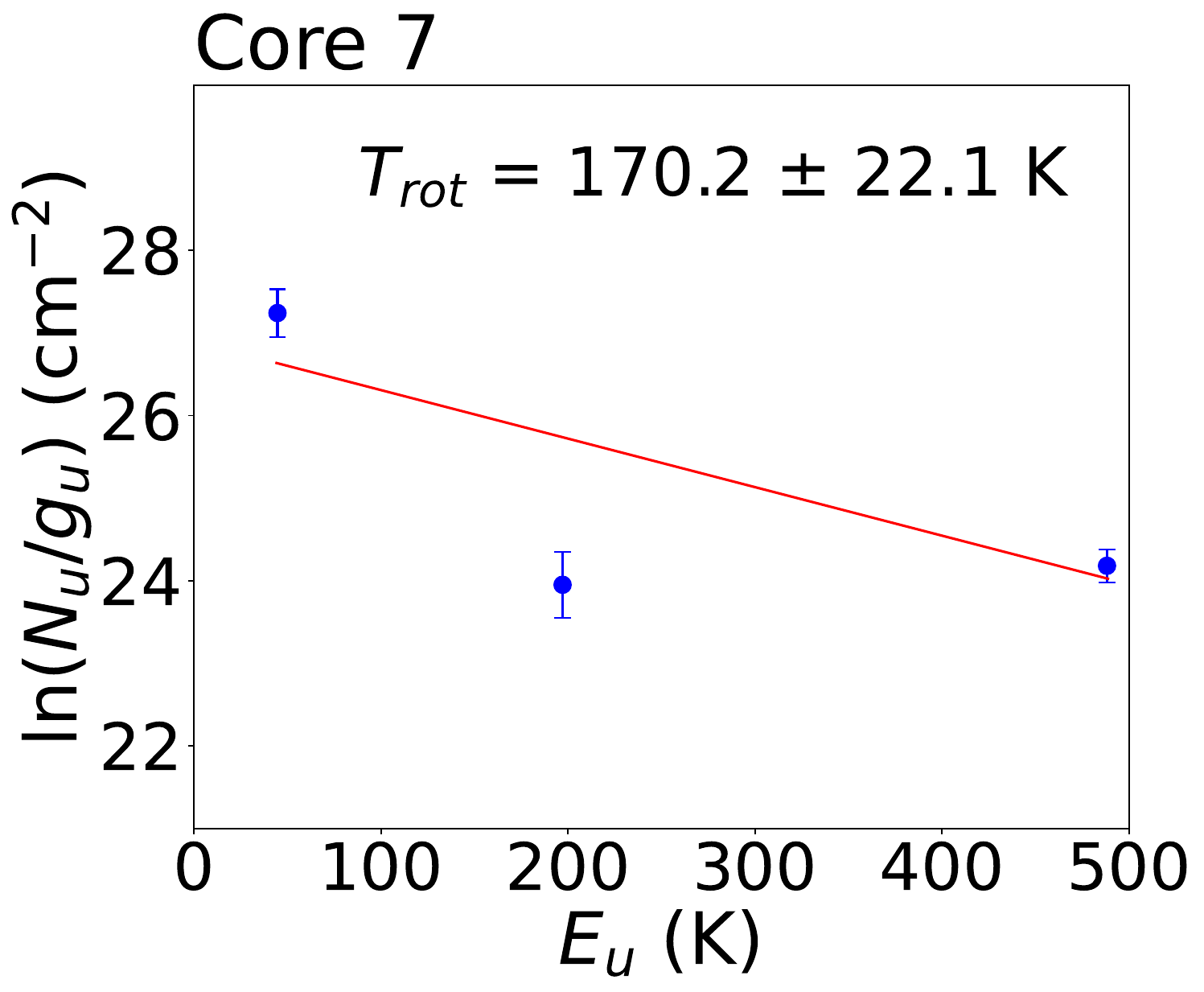}
   \includegraphics[width=0.24\linewidth]{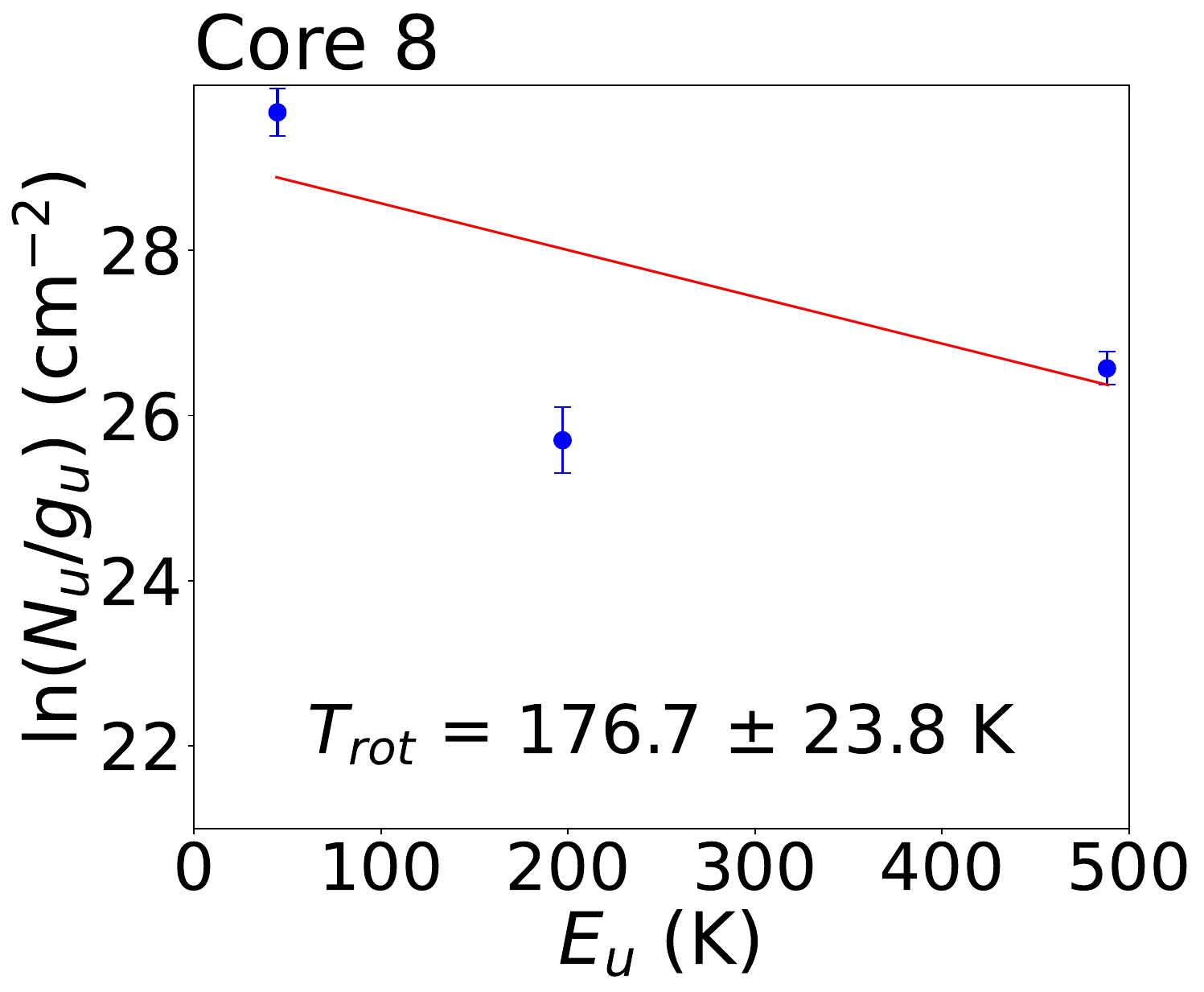}
   \includegraphics[width=0.24\linewidth]{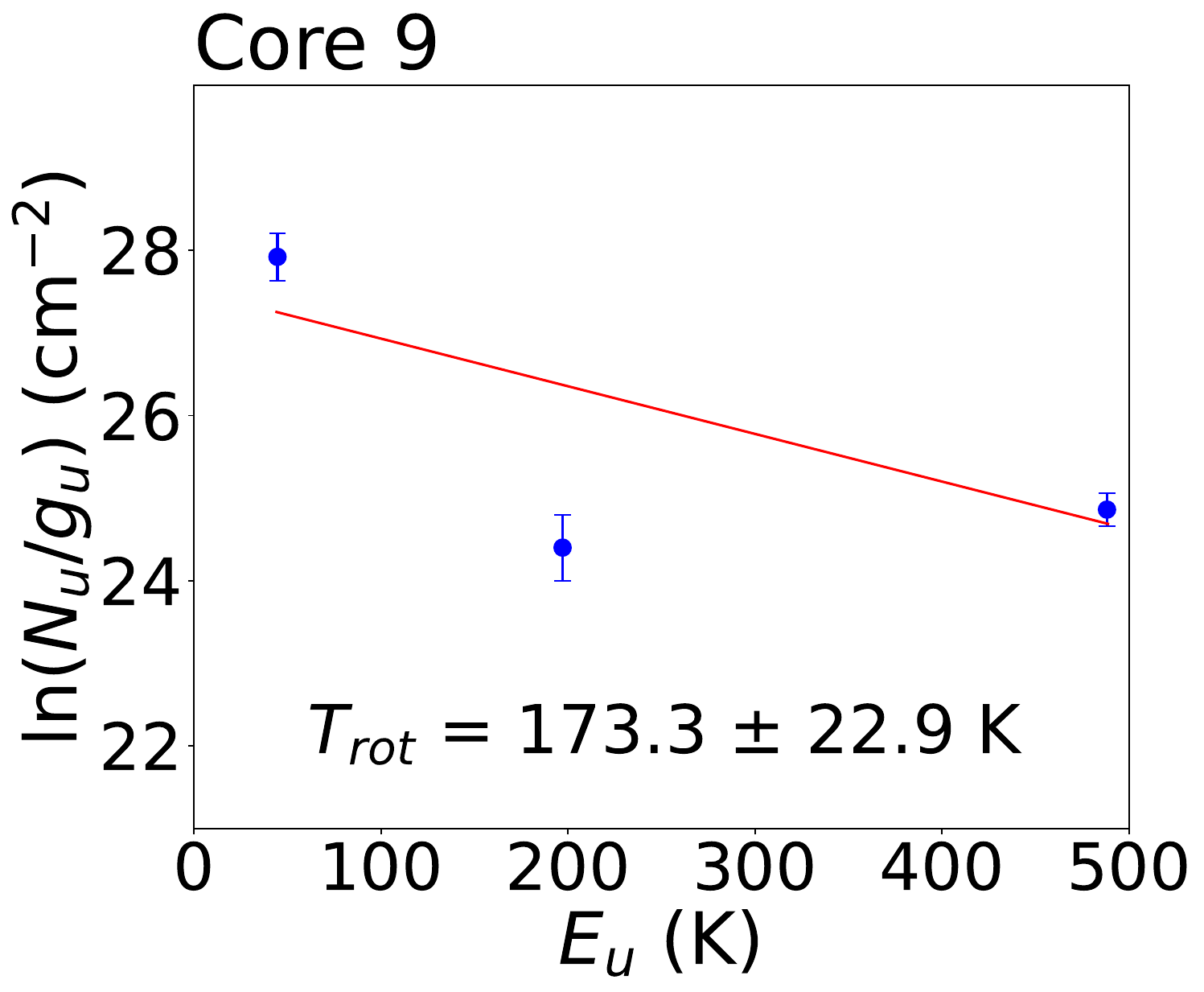}
   \includegraphics[width=0.24\linewidth]{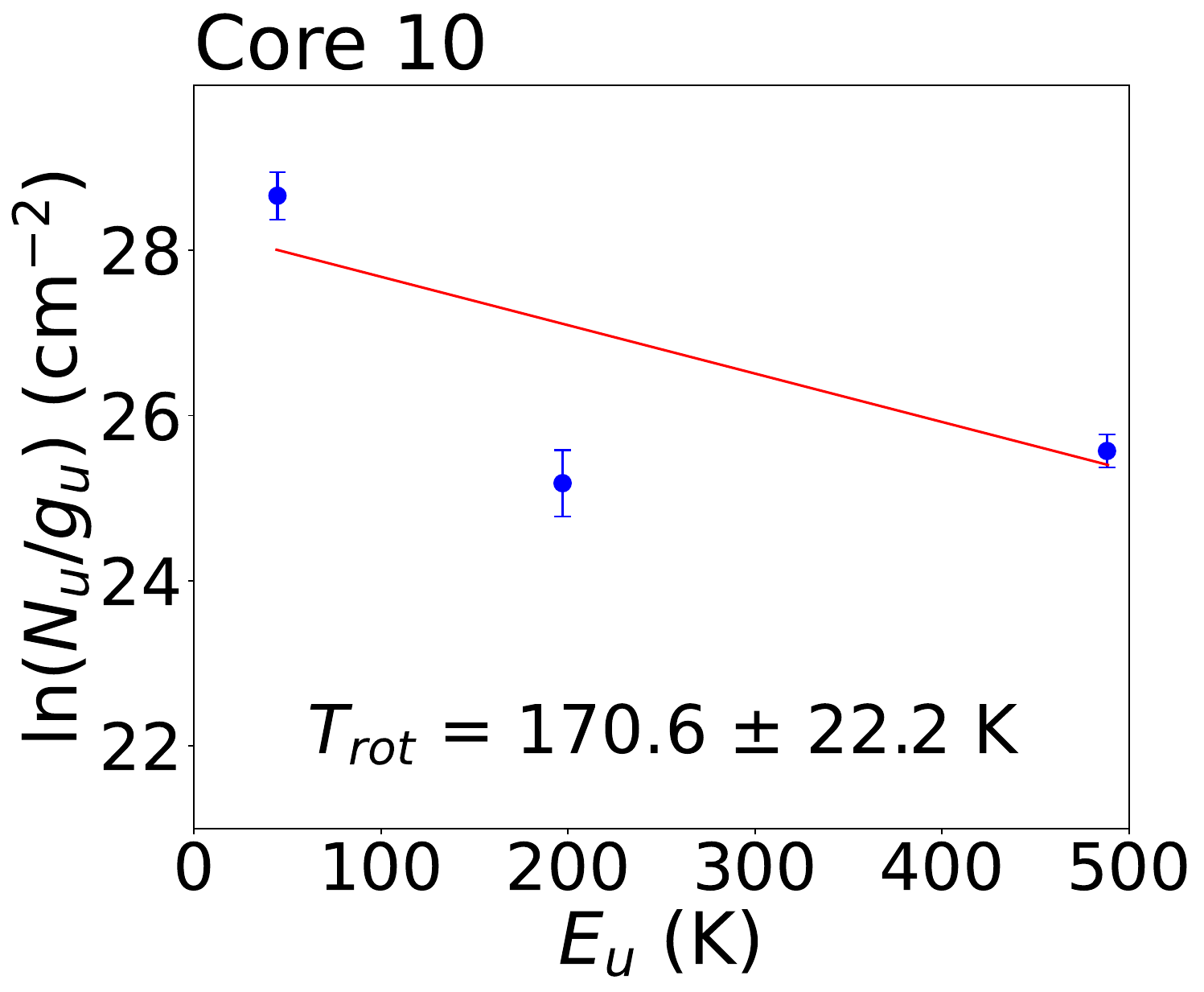}
   \includegraphics[width=0.24\linewidth]{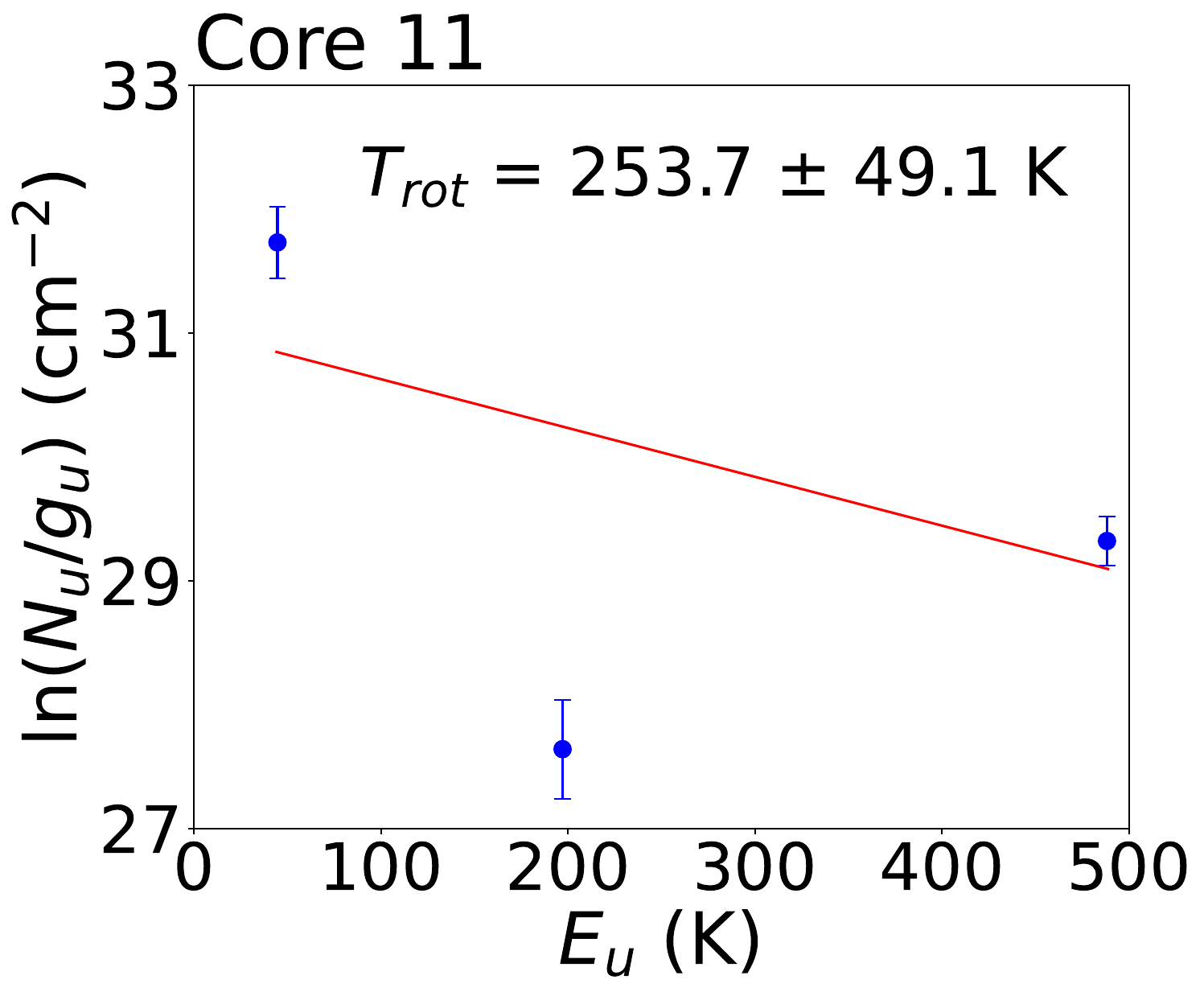}
   \includegraphics[width=0.24\linewidth]{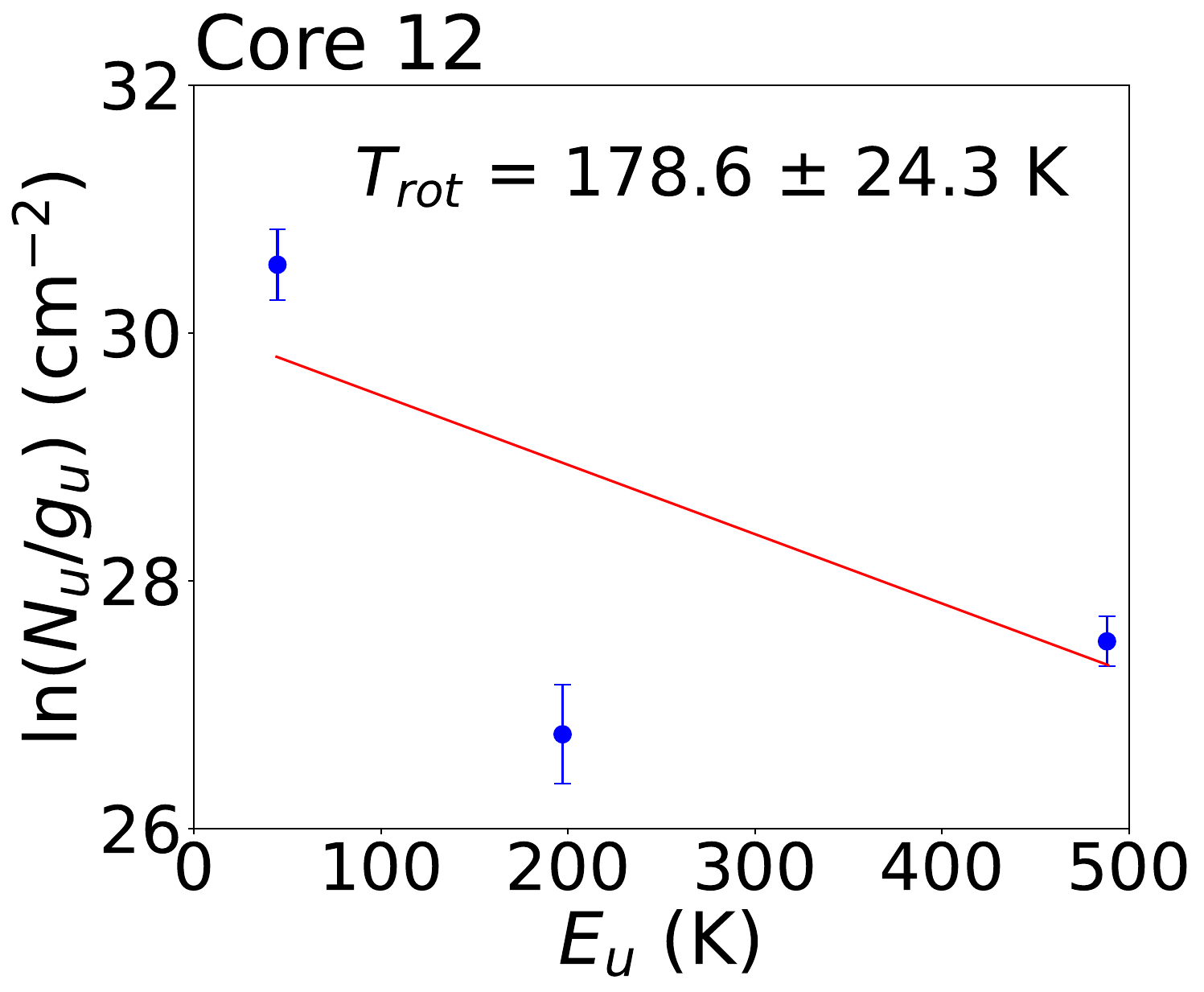}
   \includegraphics[width=0.24\linewidth]{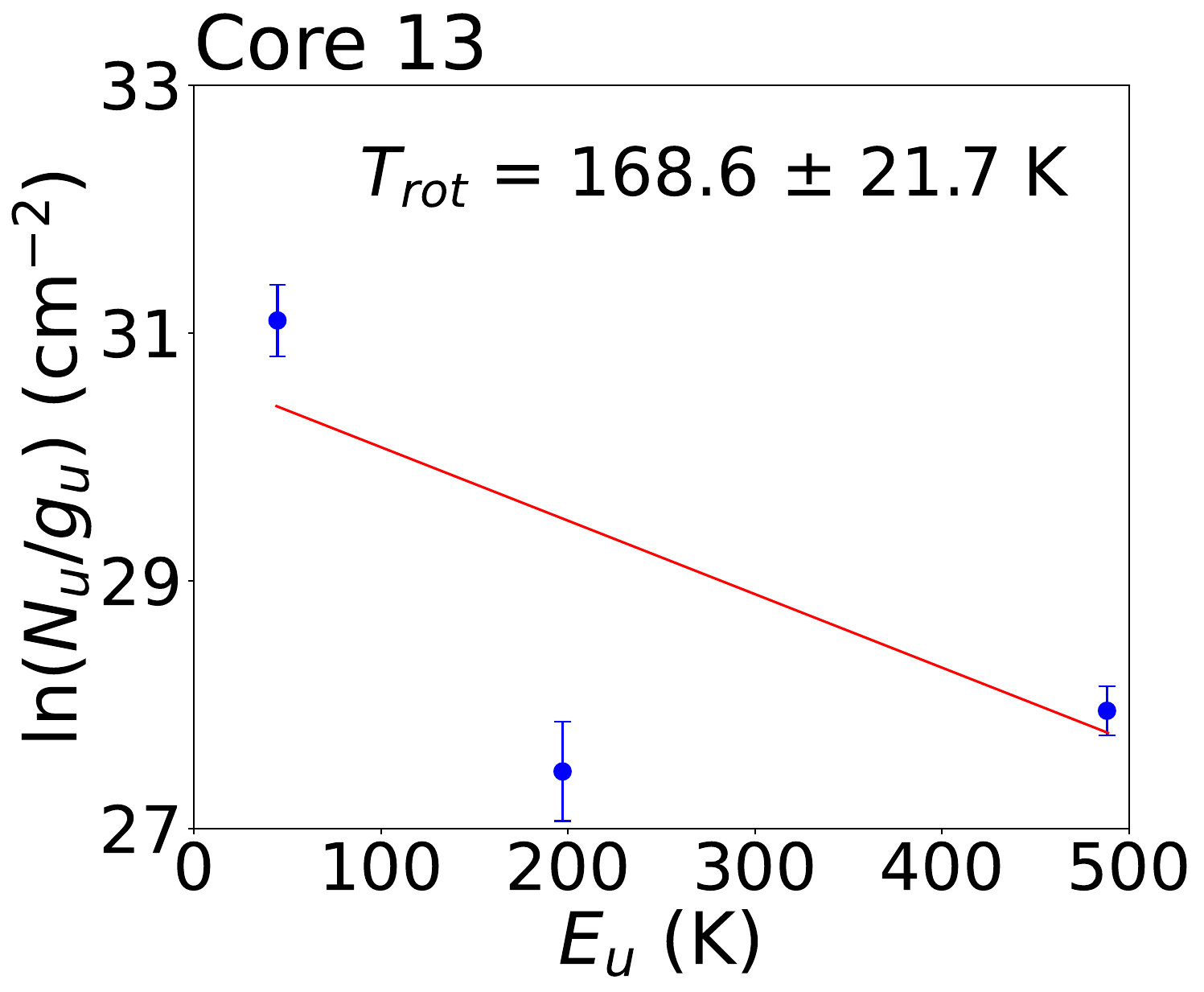}
   \includegraphics[width=0.24\linewidth]{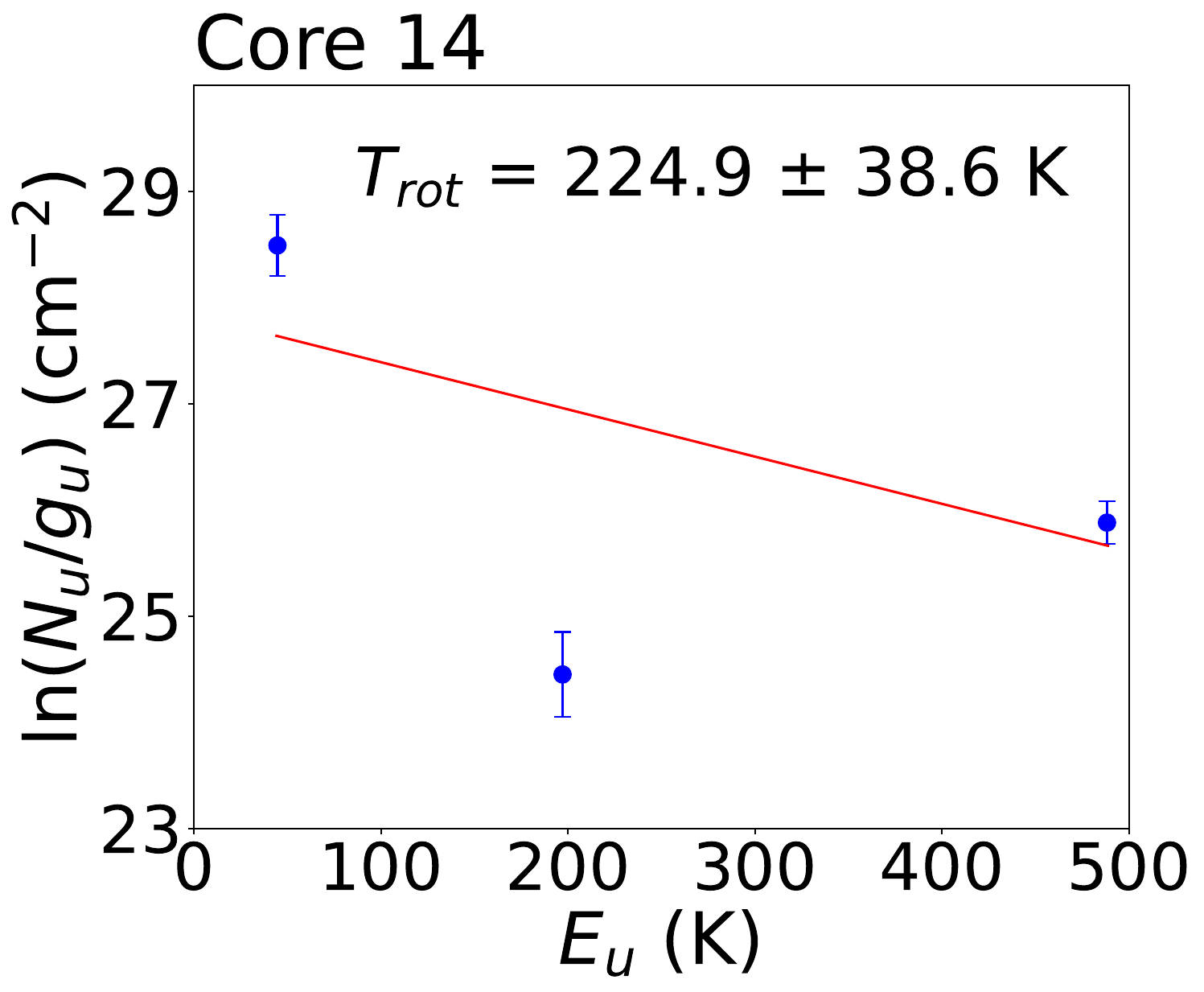}
   \includegraphics[width=0.24\linewidth]{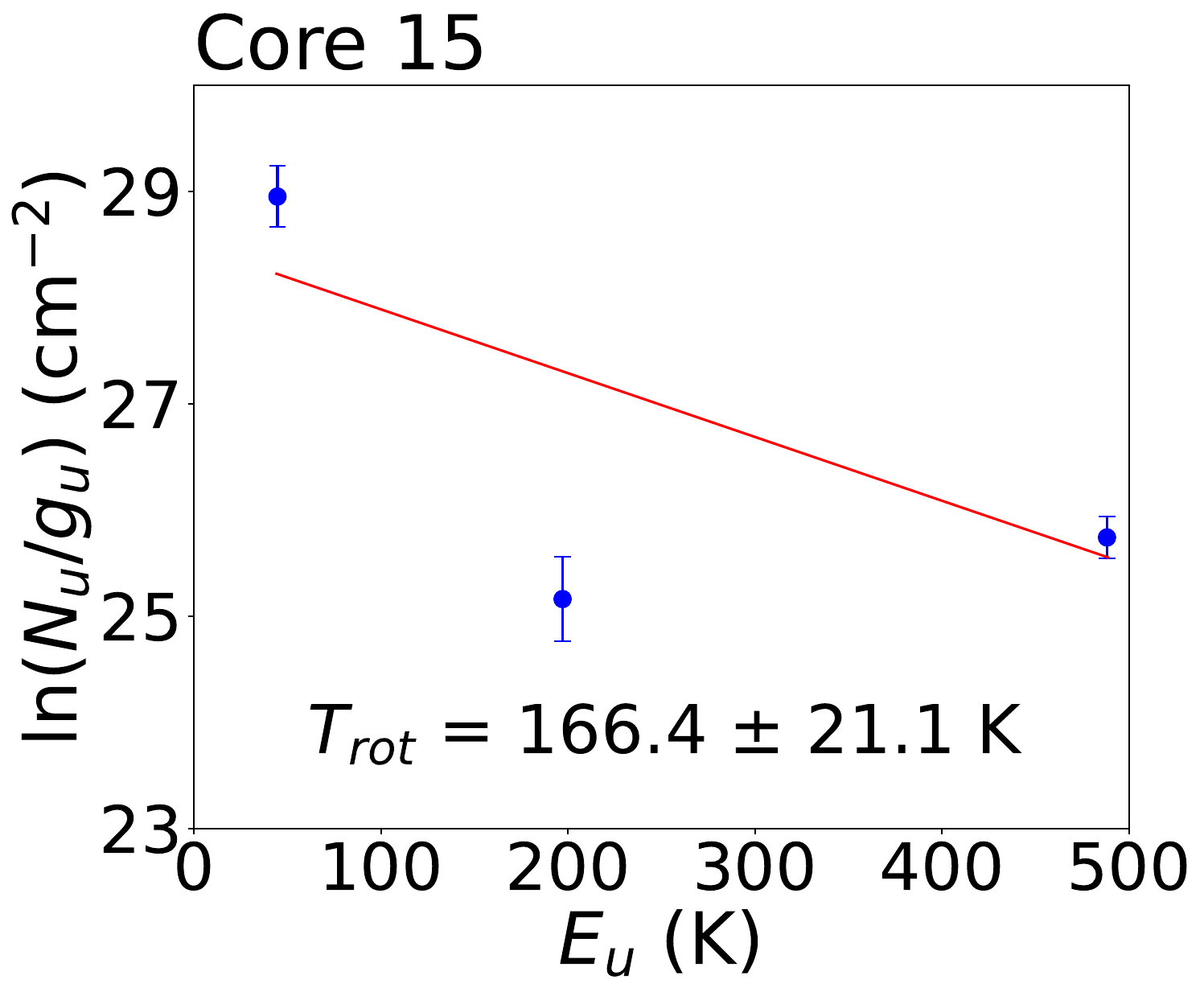}
      \includegraphics[width=0.24\linewidth]{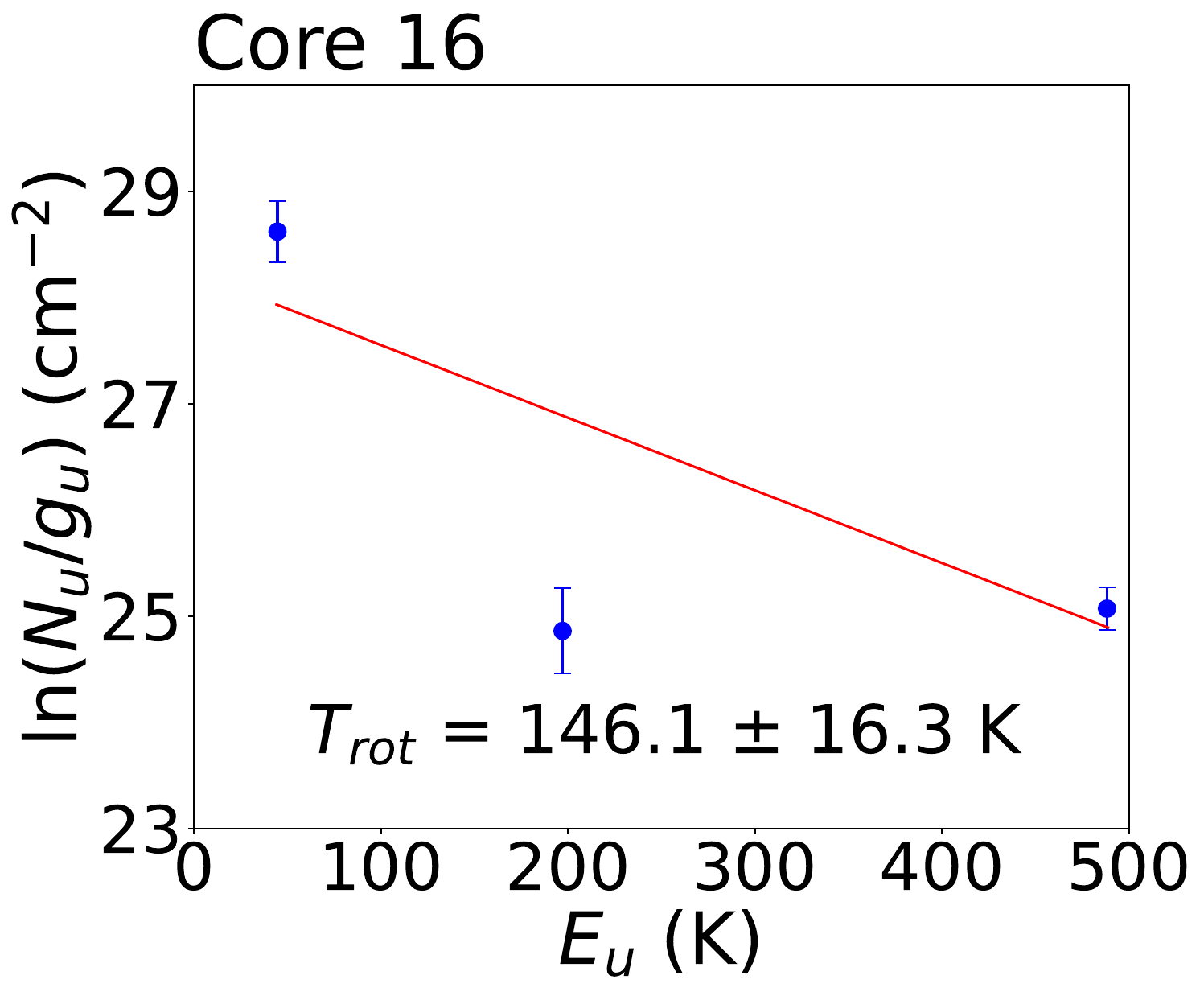}
    \caption{Rotational diagrams performed to the methanol transitions whose fluxes are presented in Table\,\ref{metanol}.}
    \label{RDs}
\end{figure*}

Table\,\ref{metanol} presents the integrated emission from the CH$_{3}$OH 2(2,1)-3(1,2)(-\,-) at 335.133 GHz ($E_{\rm u}$=44.6 K), 12(1,11)-12(0,12)(-+) at 336.865 GHz ($E_{\rm u}$=197.07 K), and 14(7,8)-15(5,9)(++) at 336.865 GHz ($E_{\rm u}$=488.21 K) lines used to estimate temperatures. 
Such integrated intensities, obtained from Gaussian fittings were used to construct rotational diagrams following \citet{goldsmith99}. By assuming LTE conditions, optically thin lines, and a beam filling factor equal to the unity, we can estimate the rotational temperatures ($ T_{\rm rot}$) from the slopes of the linear fittings in the $ln(N_{\rm u}/g_{\rm u})$ vs $E_{\rm u}$ plots displayed in Fig.\,\ref{RDs}. For more details about the procedure see \citet{martinez24} (Appendix\,C). The obtained $T_{\rm rot}$ is also included in Table\,\ref{metanol} (Col.\,5). Typical errors in the temperature are about 10\%.

\begin{table}[h]
\centering
\begin{threeparttable}
\caption{CH$_{3}$OH integrated emission lines and derived $T_{\rm rot}$.}
\label{metanol}
\tiny
\begin{tabular}{lcccccccc}
\hline
\hline
\# & $W1$ & err. & $W2$ & err. & $W3$ & err. & $T_{\rm rot}$ & err. \\ 
\hline
1  & 1.22   & 0.06  & 2.25  & 0.07  & 0.71   & 0.08   & 219.8 & 36.9 \\
2  & 5.90   & 0.28 & 7.00  & 0.37  & -      & -      & 44.7   & 6.4 \\
3  & 8.44   & 0.19  & 31.90 & 0.33  & 1.98   & 0.07   & 141.4 & 15.2 \\
4  & 1.21   & 0.07  & 3.07  & 0.18  & -      & -      & 45.0  & 6.5 \\
5  & 7.65   & 0.49  & 7.17  & 0.28  & 1.37   & 0.17   & 140.7  & 15.1\\
6  & 3.19   & 0.19  & 12.53 & 0.23  & -      & -      & 51.5  & 8.5 \\
7  & 4.68   & 0.11  & 3.16  & 0.36  & 1.70   & 0.12   & 170.2  & 22.1  \\
8  & 50.47  & 2.91  & 71.53 & 2.29  & 17.61  & 0.18   & 176.7  & 23.8 \\
9  & 9.19   & 0.73  & 20.62 & 0.42  & 3.35   & 0.13   & 173.3   & 22.9\\
10 & 19.37  & 0.24  & 44.85 & 0.37  & 6.81   & 0.09   & 170.6  & 22.2 \\
11 & 355.85 & 5.99  & 446.72& 13.29 & 248.65 & 4.20   & 253.7  & 49.1 \\
12 & 114.97 & 4.67  & 196.36& 1.59  & 42.91  & 1.40   & 178.6  & 24.3\\
13 & 188.41 & 4.43  & 373.09& 11.72 & 62.90  & 1.85   & 168.6  & 21.7\\
14 & 14.65  & 0.22  & 19.45 & 0.56  & 8.37   & 0.12   & 224.9  & 38.6 \\
15 & 22.04  & 0.86  & 37.43 & 1.08  & 6.88   & 0.17   & 166.4  & 21.1 \\
16 & 15.87  & 0.99  & 27.69 & 0.88  & 3.52   & 0.08   & 146.1  & 16.3\\
\hline
\end{tabular}
\begin{tablenotes}[para,flushleft]
  \tiny
  \item $W1$, $W2$, and $W3$ are the fluxes of the CH$_{3}$OH 2(2,1)-3(1,2)(-\,-), 12(1,11)-12(0,12)(-+), and 14(7,8)-15(5,9)(++) lines, respectively. The unity of the fluxes is Jy beam$^{-1}$ \ks~and of the $T_{\rm rot}$ is K.
\end{tablenotes}
\end{threeparttable}

\end{table}

The ${\rm T_{rot}}$ was estimated by assuming optically thin lines (e.g. \citealt{vander2000}). However, we consider the possibility of high optical depths in the used CH$_{3}$OH lines.  Thus, following \citet{goldsmith99} and \citet{purcell2006}, we compute the opacity correction factors for the $N_{\rm u}$ of both lines. We observe that in the cases that it makes sense to apply such a correction, it modifies the $N_{\rm u}$ value in a similar way for all transitions, and therefore it does not affect the $T_{\rm rot}$ estimation.

\section{Line widths}
\label{appendDeltaV}

\begin{figure*}[h]
    \centering
    \includegraphics[width=0.9\linewidth]{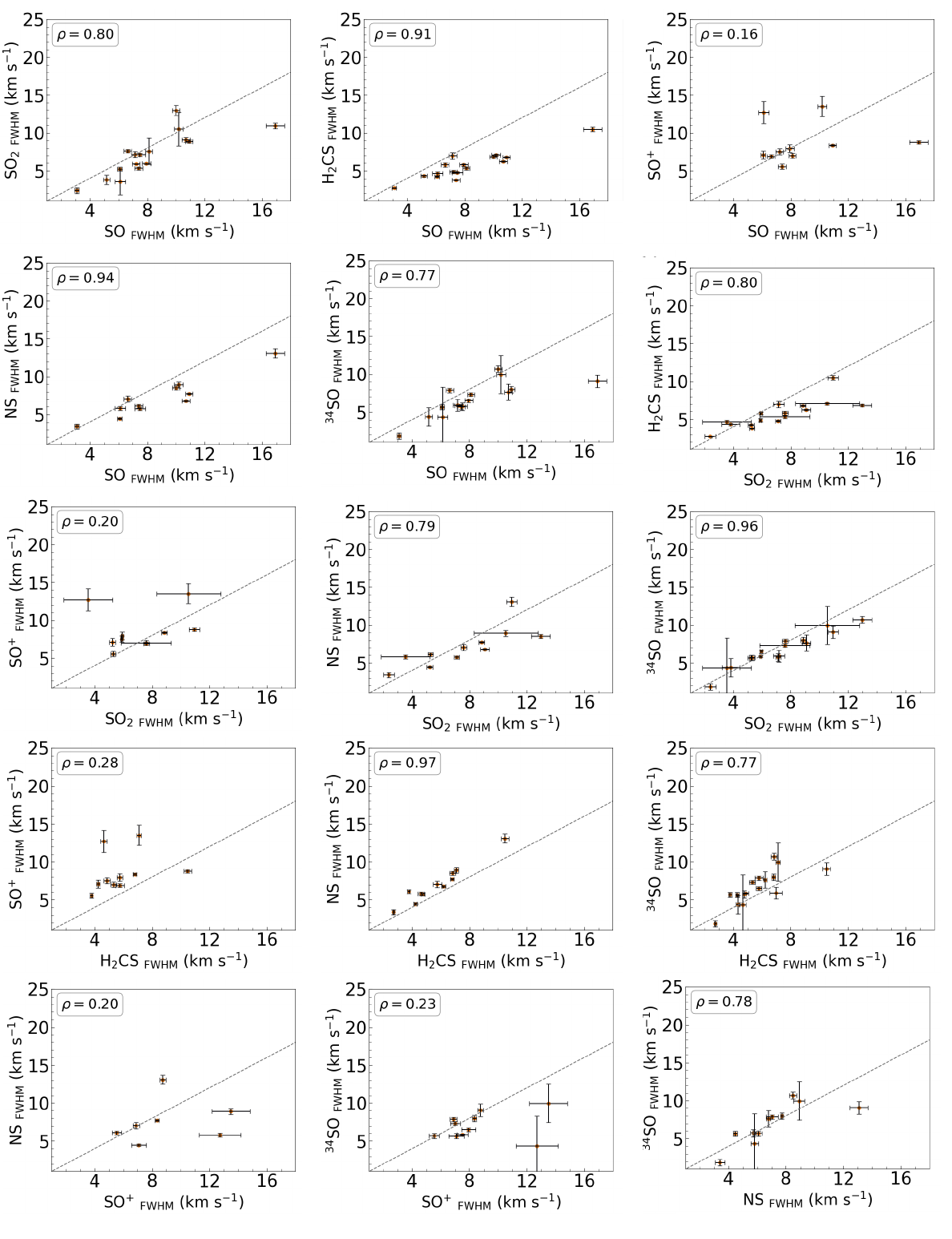}
    \caption{Comparison of the $\Delta$v FWHM of the molecular lines. 
The dashed gray line represents the y = x locus. The Pearson correlation coefficient ($\rho$) is displayed in each panel.}
    \label{deltaVfig}
\end{figure*}

Figure\,\ref{deltaVfig} present the relation between measured $\Delta$v FHWM of the sulfur-bearing
species in typical scatter plots. 

\end{appendix}
\end{document}